\documentclass[aps,preprint,amsmath,amssymb,amsfonts]{revtex4}
\usepackage{epsfig}
\usepackage{graphicx}
\usepackage{subfigure}
\usepackage{dcolumn}
\usepackage{bm}
\usepackage{amsthm}
\usepackage{enumitem}
\usepackage{slashed}
\usepackage{braket}
\usepackage{amsmath}
\usepackage{mathtools}

\makeatletter
\def\l@subsubsection#1#2{}
\makeatother

\newcommand{\del}{\partial}

\newcommand{\diag}{\operatorname{diag}}

\newcommand{\bbR}{\mathbb{R}}

\newcommand{\calD}{\mathcal{D}}

\newcommand{\calF}{\mathcal{F}}
\newcommand{\calG}{\mathcal{G}}

\newcommand{\calO}{\mathcal{O}}

\newcommand{\calT}{\mathcal{T}}

\begin{document}

\title{Higher-spin gravity's ``string'': new gauge and proof of holographic duality for the linearized Didenko-Vasiliev solution}

\author{Vyacheslav Lysov}
\email{vyacheslav.lysov@oist.jp}
\affiliation{Okinawa Institute of Science and Technology, 1919-1 Tancha, Onna-son, Okinawa 904-0495, Japan}
\author{Yasha Neiman}
\email{yashula@icloud.com}
\affiliation{Okinawa Institute of Science and Technology, 1919-1 Tancha, Onna-son, Okinawa 904-0495, Japan}

\date{\today}

\begin{abstract}
We consider type-A higher-spin gravity in AdS\textsubscript{4}, holographically dual to a free $U(N)$ vector model on the boundary. We study the linearized version of the Didenko-Vasiliev ``BPS black hole'', which we view as this theory's equivalent of the fundamental string. The Didenko-Vasiliev solution consists of gauge fields of all spins generated by a particle-like source along a bulk geodesic, and is holographically dual to a bilocal boundary operator at the geodesic's endpoints. Our first main result is a new gauge for this solution, which makes manifest its behavior under the boundary field equation. It can be viewed as an AdS uplift of flat spacetime's de Donder gauge, but is not de Donder in AdS. To our knowledge, this gauge is novel even in the spin-2 sector, and thus provides a new expression for the linearized gravitational field of a massive point particle in (A)dS\textsubscript{4}. Our second main result is a proof of the holographic duality between the mutual bulk action of two Didenko-Vasiliev solutions and the CFT correlator of two boundary bilocals. As an intermediate step, we show that in a bilocal$\rightarrow$local limit, the Didenko-Vasiliev solution reproduces the standard boundary-bulk propagators of all spins. We work in the ``metric-like'' language of Fronsdal fields, and use the embedding-space formalism. 
\end{abstract}

\maketitle
\tableofcontents
\newpage

\section{Introduction} \label{sec:intro}

\subsection{Why study higher-spin gravity}

Higher-spin gravity is the conjectured interacting theory of an infinite tower of massless fields with increasing spins. It is simultaneously a larger version of supergravity, and a smaller version of string theory. Unlike supergravity, it is equally happy with both signs of the cosmological constant $\Lambda$ (but apparently requires $\Lambda\neq 0$). Unlike string theory, it's native to 4 spacetime dimensions, with a tight relationship to twistor theory. Like string theory, HS gravity admits an AdS/CFT holographic description. While the typical holographic duals of string theory are matrix-like gauge theories \cite{Maldacena:1997re,Gubser:1998bc,Witten:1998qj,Aharony:1999ti}, the holographic duals of HS gravity are vector models \cite{Klebanov:2002ja,Sezgin:2002rt,Sezgin:2003pt,Giombi:2012ms}. In some versions (with the HS fields decorated by color \& a high degree of supersymmetry), there is holographic evidence that HS gravity at strong coupling is dual to string theory itself \cite{Chang:2012kt,Honda:2017nku}, and thus in particular should have a supergravity limit. On the other hand, in other versions (minimal field content, no supersymmetry), it appears that the AdS/CFT duality can be continued to $\Lambda>0$ \cite{Anninos:2011ui}, which makes this the only known model of de Sitter holography with 4 (or more) bulk dimensions. The main disappointment is that the unadorned versions of HS theory which seem to work in de Sitter space are not the ones dual to string theory. In particular, they don't seem to have a GR limit, or any separation of scales that would decouple the higher spins $s>2$ from the standard ones $s\leq 2$ of GR+matter. However, beggars cannot be choosers: when a theory promises to be a working model of dS/CFT, it's our duty to understand it. 

In this work, we consider HS theory in 4 dimensions, with one field for each integer spin, and with all the fields parity-even: the so-called type-A theory. For simplicity, we consider Euclidean spacetime signature and $\Lambda<0$, i.e. we work in Euclidean AdS\textsubscript{4}. We choose the simplest boundary conditions at infinity, which preserve global HS symmetry. The holographic dual in this case is a free $U(N)$ vector model of spin-0 fields in 3 dimensions. The version relevant for dS/CFT is obtained from this theory by simply truncating to even spins, which changes the boundary color group to $O(2N)$ (for AdS/CFT) or $Sp(2N)$ (for dS/CFT).

\subsection{Why study the Didenko-Vasiliev solution}

A major challenge for higher-spin theory is its apparent non-locality \cite{Sleight:2017pcz}, which arises when one attempts to express the 4-point boundary correlator in the standard language of bulk Witten diagrams (most of the relevant calculation is given in \cite{Bekaert:2015tva}). This result was foreshadowed some years previously \cite{Fotopoulos:2010ay,Taronna:2011kt}. We should also note an earlier challenge to the theory's locality \cite{Boulanger:2015ova}, where it was found that Vasiliev's original formulation \cite{Vasiliev:1990en,Vasiliev:1995dn,Vasiliev:1999ba} leads to non-local results already for the \emph{cubic} vertex. This has since been resolved \cite{Didenko:2018fgx}, inspiring a research program to recover locality at quartic order and higher \cite{Gelfond:2019tac,Didenko:2019xzz}, which hasn't succeeded so far.

Now, it was conjectured in \cite{Fotopoulos:2010ay} that, to restore locality at quartic order, one may need to add some additional degrees of freedom. To identify a good candidate for such additional degrees of freedom, we propose to take a cue from string theory. Like HS gravity, string theory can be thought of as an interacting theory of infinitely many fields -- the modes of the string. Outside the supergravity limit (where only a finite number of light fields survive), this theory is non-local. On the other hand, the theory is perfectly local once it is understood in terms of the string worldsheet. The worldsheet isn't quite adding new degrees of freedom to the field picture: the fields constitute a complete mode expansion of it. However, it does offer a ``non-perturbative completion'' to the tower of fields, which re-frames the locality discussion. 

What, then, is HS gravity's analogue of string theory's fundamental string? To help answer this question, let us recall some further aspects of the string worldsheet. The first is that the fundamental string, along with other extended objects of string theory (D-branes, M-branes, NS5-branes) can be identified with a BPS black brane solution of supergravity (see e.g. \cite{Schwarz:1996bh,Blumenhagen:2013fgp}). Now, the Vasiliev equations of HS gravity have their own version of a BPS solution -- the Didenko-Vasiliev ``black hole'' \cite{Didenko:2009td}. Might this be HS gravity's analogue of the fundamental string? We argue that the answer is yes.

To support this conclusion, consider yet another potential route for ``discovering'' the string. In AdS/CFT, the field spectrum of the string becomes the spectrum of primary single-trace operators in the boundary gauge theory. Now, within the boundary theory, one can view these local single-trace operators as a Taylor expansion of a Wilson loop (expanded around vanishing loop area). One can then discover the string -- the ``non-perturbative completion'' of the tower of bulk fields, as the holographic dual of the Wilson loop -- the ``non-perturbative completion'' of the tower of local boundary operators \cite{Rey:1998ik,Maldacena:1998im}. In HS holography with a free vector model on the boundary, the tower of primary local boundary operators is a tower of HS currents. The non-local object that contains and ``completes'' this tower is the \emph{bilocal product} \cite{Das:2003vw,Douglas:2010rc,Das:2012dt} of the vector models' fields at two points. The bilocal also appears naturally as an intermediate object in calculations of $n$-point correlators in the free boundary CFT: when we Wick-contract a chain of HS currents within the correlator, the result is a bilocal \cite{Didenko:2012tv,Neiman:2017mel}. It is then natural to expect that the bilocal should play a key role in the bulk interactions dual to these boundary correlators. 

Now, the bulk dual of boundary bilocals was computed in \cite{Neiman:2017mel}, and identified in \cite{David:2020fea} as the \emph{linearized version \cite{Didenko:2008va} of the Didenko-Vasiliev solution}, with the precise BPS-like pattern of HS charges dictated by the master-field description in \cite{Didenko:2009td}. We thus view the Didenko-Vasiliev solution as key to illuminating the structure of HS gravity, including its relationship with locality. 

As an aside, we note that the Didenko-Vasiliev solution belongs to an infinite-parameter family of solutions constructed in \cite{Iazeolla:2011cb}, which have arbitrary charges with respect to the HS fields of different spins. Conversely, one can consider the boundary bilocal as ``merely'' a particular choice of generating function for HS currents, again with a particular choice of relative normalizations for the different spins. These potential freedoms of normalization on the bulk and boundary are related to one another. In the duality between the boundary bilocal and the bulk Didenko-Vasiliev solution, both freedoms are fixed in the most natural way.

\subsection{Summary and structure of the paper}

In this paper, we present some basic results concerning the linearized version \cite{Didenko:2008va} of the Didenko-Vasiliev solution, building on previous work in \cite{Neiman:2017mel,David:2020ptn,David:2020fea}. We work in an embedding-space formalism, using the language of Fronsdal fields \cite{Fronsdal:1978rb,Fronsdal:1978vb}. The formalism, along with the Didenko-Vasiliev solution itself, is reviewed in section \ref{sec:preliminaries}. In three sentences, what the reader should take away from these preliminaries is:

\emph{The linearized Didenko-Vasiliev ``black hole'' is a solution to Fronsdal's linearized HS field equations, in a traceless gauge, with a particle-like source concentrated on a bulk geodesic. The geodesic carries charges of all spins, which follow a BPS-like proportionality pattern. This solution is holographically dual to a bilocal boundary operator, supported on the geodesic's endpoints.}

Section \ref{sec:preliminaries} also contains some new geometric formulas, in particular regarding bulk geodesics and their associated delta functions. 

The paper's central result is an analytic proof of a holographic relationship that was conjectured and checked numerically in \cite{David:2020fea}: the equality between the mutual bulk action of two Didenko-Vasiliev solutions, and the CFT correlator of two boundary bilocals. This proof is presented in section \ref{sec:analytic_2_point}, using constructions and results from the other sections. Let us detail these in turn.

Our first result (section \ref{sec:local_limit}) concerns the bilocal$\rightarrow$local limit on the boundary. We show explicitly how the Didenko-Vasiliev solution (the holographic dual to a boundary bilocal) reproduces in an appropriate limit to ordinary boundary-bulk propagators (the holographic duals to local boundary currents). This extends to the level of gauge potentials our previous result \cite{David:2020ptn} for field strengths, and allows us to prove a limiting case of the general holographic statement of section \ref{sec:analytic_2_point}. Our next result (section \ref{sec:Box}) concerns the application of the boundary field equation, i.e. the boundary conformal Laplacian, to the Didenko-Vasiliev solution. For correlation functions of boundary bilocals, this Laplacian vanishes away from coincident points. Holography (along with the field-strength results of \cite{Neiman:2017mel}) leads us to expect a similar result for the Didenko-Vasiliev solution. Accordingly, we find that its boundary Laplacian can be expressed as the sum of (a) a pure-gauge field and (b) a delta-function term with support on the associated bulk geodesic. Removing the pure-gauge piece and undoing the boundary Laplacian, we end up with a new gauge for the Didenko-Vasiliev solution (section \ref{sec:Phi}). This new gauge doesn't satisfy any standard bulk gauge condition, but instead has an especially simple relationship with one of the \emph{boundary} endpoints of the geodesic. To our knowledge, this gauge is new even for spin 2, i.e. for the gravitational field of a point particle in (A)dS\textsubscript{4} in linearized GR. In a flat-spacetime limit, our new gauge becomes de Donder. 

Our main proof in section \ref{sec:analytic_2_point} combines a calculation in the new gauge of section \ref{sec:Phi}, a bound on the effect of the gauge transformation from section \ref{sec:Box}, and the local-limit result of section \ref{sec:local_limit}. Finally, in section \ref{sec:double_Box}, we calculate for completeness the \emph{double} Laplacian of the Didenko-Vasiliev solution (in the new gauge), with respect to \emph{both} boundary endpoints.

\section{Preliminaries} \label{sec:preliminaries}

\subsection{Bulk geometry} \label{sec:preliminaries:bulk_geometry}

In this paper, we will work in an embedding-space formalism. Thus, we describe Euclidean AdS\textsubscript{4} as the hyperboloid of unit timelike radius within 5d flat spacetime $\bbR^{1,4}$:
\begin{align}
 EAdS_4 = \left\{x^\mu\in\bbR^{1,4}\, |\, x_\mu x^\mu = -1, \ x^0 > 0 \right\} \ . \label{eq:EAdS}
\end{align}
Here, indices $(\mu,\nu,\dots)$ are 5-dimensional, and are raised and lowered with the Minkowski metric $\eta_{\mu\nu} = \diag(-1,1,1,1,1)$. 4d vectors at a point $x^\mu \in EAdS_4$ are simply 5d vectors $v^\mu$ that satisfy $v\cdot x\equiv v_\mu x^\mu = 0$. Covariant derivatives in $EAdS_4$ are simply flat $\bbR^{1,4}$ derivatives, followed by a projection of all indices back into the $EAdS_4$ tangent space:
\begin{align}
 \nabla_\mu v_\nu = P_\mu^\rho(x) P_\nu^\sigma(x) \frac{\del v_\sigma}{\del x^\rho} \ ; \quad P_\mu^\nu(x) \equiv \delta_\mu^\nu + x_\mu x^\nu \ . \label{eq:nabla}
\end{align}
With lowered indices, the projector $P_\mu^\nu(x)$ becomes the 4d metric of $EAdS_4$ at $x$:
\begin{align}
 g_{\mu\nu}(x) = \eta_{\mu\nu} + x_\mu x_\nu \ . \label{eq:g}
\end{align}
Our use of different letters for $P_\mu^\nu$ and $g_{\mu\nu}$ is purely cosmetic.

Since HS fields carry many symmetrized tensor indices, it is convenient to package them as functions of an auxiliary ``polarization vector'' $u^\mu\in\bbR^{1,4}$. Thus, we encode a rank-$p$ symmetric tensor by a function of the form:
\begin{align}
 f(x,u) = \frac{1}{p!}\,u^{\mu_1}\dots u^{\mu_p}f_{\mu_1\dots\mu_p}(x) \ . \label{eq:generating_function}
\end{align}
The tensor rank of $f_{\mu_1\dots\mu_p}$, and the fact that it's tangential to the $EAdS_4$ hyperboloid, can be expressed as constraints on $f(x,u)$: 
\begin{align}
 (u\cdot\del_u) f = pf \ ; \quad (x\cdot\del_u) f = 0 \ ,
\end{align}
where $\del_u^\mu$ denotes the flat $\bbR^{1,4}$ derivative with respect to $u^\mu$. Tracing a pair of indices on $f_{\mu_1\dots\mu_p}$ is encoded by acting on $f(x,u)$ with the operator $\del_u\cdot\del_u$. A symmetrized product with the $EAdS_4$ metric \eqref{eq:g} is encoded as multiplication by $g_{\mu\nu} u^\mu u^\nu$. The most natural differential operators on the space of totally symmetric tensors are the divergence $\del_u\cdot\nabla$, the symmetrized gradient $u\cdot\nabla$ and the Laplacian $\nabla\cdot\nabla$, where $\nabla_\mu$ is the $EAdS_4$ covariant derivative \eqref{eq:nabla}.

Given a $EAdS_4$ vector $\hat t^\mu$ at $x$, the \emph{traceless part} of the symmetric tensor $\hat t^{\mu_1}\dots\hat t^{\mu_s}$ is encoded by the function:
\begin{align}
 \begin{split}
   \calT^{(p)}(x,\hat t,u) &= \frac{(\hat t\cdot u)^p}{p!} - \text{traces} = \frac{1}{p!}\sum_{n=0}^{\lfloor p/2 \rfloor} \binom{p-n}{n}\left(-\frac{1}{4}(\hat t\cdot\hat t)(g_{\mu\nu}(x)u^\mu u^\nu) \right)^n (\hat t\cdot u)^{p-2n} \\
     &= \frac{1}{2^p p!}\sum_{n=0}^{\lfloor p/2 \rfloor} \binom{p+1}{2n+1} \big({-(\hat t\cdot\hat t)}(q_{\mu\nu}(x,\hat t)u^\mu u^\nu) \big)^n (\hat t\cdot u)^{p-2n} \ ,
 \end{split} \label{eq:T}
\end{align}
where ``${}- \text{traces}$'' refers to subtracting $\sim g_{\mu\nu} u^\mu u^\nu$ pieces to make the result traceless. In the second line of \eqref{eq:T}, we introduced the 3d metric $q_{\mu\nu} = g_{\mu\nu} - \frac{\hat t_\mu\hat t_\nu}{\hat t\cdot\hat t}$ of the subspace orthogonal to both $x^\mu$ and $\hat t^\mu$.

\subsection{Fronsdal fields in the bulk} \label{sec:preliminaries:Fronsdal}

Let us review the form of Fronsdal's field equations for linearized HS fields \cite{Fronsdal:1978vb} in the above framework. In Fronsdal's formalism, a spin-$s$ field (more precisely, gauge potential) is a totally symmetric rank-$s$ tensor with vanishing \emph{double trace}. This can be encoded by a scalar function $h^{(s)}(x,u)$, as in \eqref{eq:generating_function}, subject to the constraints:
\begin{align}
 (u\cdot \del_u)h^{(s)} = s h^{(s)} \ ; \quad (x\cdot\del_u)h^{(s)} = 0 \ ; \quad (\del_u\cdot\del_u)^2 h^{(s)} = 0 \ . \label{eq:h_tensor_properties}
\end{align}
Gauge transformations take the form:
\begin{align}
h^{(s)} \ \rightarrow \ h^{(s)} + (u\cdot\nabla)\Lambda^{(s)} \ , \label{eq:gauge_transformation}
\end{align}
where $\Lambda^{(s)}(x,u)$ encodes a rank-$(s-1)$ totally symmetric, traceless gauge parameter:
\begin{align}
 (u\cdot \del_u)\Lambda^{(s)} = (s-1)\Lambda^{(s)} \ ; \quad (x\cdot\del_u)\Lambda^{(s)} = 0 \ ; \quad (\del_u\cdot\del_u)\Lambda^{(s)} = 0 \ . \label{eq:Lambda_tensor_properties}
\end{align}
Out of the field $h^{(s)}$, we can construct a gauge-invariant curvature, which generalizes the $s=2$ linearized Ricci tensor to all spins. This is the Fronsdal tensor $\calF h^{(s)}$, where the Fronsdal operator $\calF$ is given by:
\begin{align}
 \calF = -\nabla\cdot\nabla + s^2 - 2s - 2 + (u\cdot\nabla)(\del_u\cdot\nabla) - \left(\frac{1}{2}(u\cdot\nabla)^2 - g_{\mu\nu}u^\mu u^\nu\right)(\del_u\cdot\del_u) \ . \label{eq:Fronsdal}
\end{align}
$\calF$ is a second-order differential operator with respect to $x$. The Fronsdal tensor $\calF h^{(s)}$ has the same tensor properties \eqref{eq:h_tensor_properties} as the potential $h^{(s)}$. In analogy with GR, we can rearrange the trace of $\calF h^{(s)}$ to obtain the Einstein tensor:
\begin{align}
 \calG h^{(s)} = \left(1 - \frac{1}{4}(g_{\mu\nu}u^\mu u^\nu)(\del_u\cdot\del_u) \right) \calF h^{(s)} \ . \label{eq:Einstein}
\end{align}
This has the same tensor properties \eqref{eq:h_tensor_properties}, but also satisfies a conservation law of the form:
\begin{align}
 (\del_u\cdot\nabla)\calG h^{(s)} = (g_{\mu\nu}u^\mu u^\nu)(\dots) \ , \label{eq:Einstein_conservation}
\end{align}
i.e. the $EAdS_4$ divergence of $\calG h^{(s)}$ vanishes up to trace terms. This allows us to write a gauge-invariant action for linearized HS fields:
\begin{align}
 S = s!\int_{EAdS_4} d^4x\,h^{(s)}(x,\del_u)\left(\frac{1}{2}\calG h^{(s)}(x,u) - J^{(s)}(x,u) \right) \ , \label{eq:Fronsdal_action}
\end{align}
with $J^{(s)}(x,u)$ an external HS current, which must be conserved in the same sense \eqref{eq:Einstein_conservation} as $\calG h^{(s)}$. Note that in general, the action \eqref{eq:Fronsdal_action} requires also a boundary term, which couples the electric and magnetic boundary data of $h^{(s)}$. In the standard use-case of computing the mutual action of two boundary-bulk propagators, this boundary term is essential. However, in our present work with the Didenko-Vasiliev solution, we will not need to consider it. Away from the boundary endpoints of the geodesic carrying the solution's HS charges, this can be justified rigorously, since the solution's magnetic boundary data vanishes \cite{Neiman:2017mel,David:2020fea}. \emph{At} the endpoints, the free-field asymptotic analysis doesn't apply, so one can imagine some, perhaps non-standard, boundary terms; however, we observe empirically that our main result \eqref{eq:bilocal_2_point_raw} for the action, as guessed and numerically tested in \cite{David:2020fea} and proved below in section \ref{sec:analytic_2_point}, holds without any such terms. In particular, in the limit \eqref{eq:local_limit} below where the Didenko-Vasiliev solution \emph{does} turn into a boundary-bulk propagator, the bulk action \eqref{eq:Fronsdal_action} limits correctly into the standard boundary term.

The field equations for the action \eqref{eq:Fronsdal_action} read simply:
\begin{align}
 \calG h^{(s)}(x,u) = J^{(s)}(x,u) \ , \label{eq:Fronsdal_equation}
\end{align}
and the action \eqref{eq:Fronsdal_action} evaluates on-shell to:
\begin{align}
 S = -\frac{s!}{2}\int_{EAdS_4} d^4x\,h^{(s)}(x,\del_u) J^{(s)}(x,u) = -\frac{s!}{2}\int_{EAdS_4} d^4x\,h^{(s)}(x,\del_u)\,\calG h^{(s)}(x,u) \ . \label{eq:on_shell_action}
\end{align}
The formalism is simplified significantly in a \emph{traceless gauge} (which can also be viewed as a framework in its own right \cite{Skvortsov:2007kz,Campoleoni:2012th}). In this gauge, the double-traceless condition $(\del_u\cdot\del_u)^2 h^{(s)} = 0$ is strengthened into just tracelessness $(\del_u\cdot\del_u) h^{(s)} = 0$. The remaining gauge freedom is parameterized by \eqref{eq:gauge_transformation}-\eqref{eq:Lambda_tensor_properties}, with the further constraint:
\begin{align}
 (\del_u\cdot\nabla)\Lambda^{(s)} = 0 \ . \label{eq:div_Lambda}
\end{align}
In this gauge, the Fronsdal operator \eqref{eq:Fronsdal} simplifies into:
\begin{align}
 \calF = -\nabla\cdot\nabla + s^2 - 2s - 2 + (u\cdot\nabla)(\del_u\cdot\nabla) \ . \label{eq:Fronsdal_traceless}
\end{align}
In particular, the linearized Didenko-Vasiliev solution of \cite{Didenko:2008va} is given in a traceless gauge. 

Outside of traceless gauge, an occasionally useful quantity is the \emph{traceless part of the divergence $(\del_u\cdot\nabla)h^{(s)}$}. This is known as the de Donder tensor $\calD h^{(s)}$, where $\calD$ is the de Donder operator:
\begin{align}
 \calD = \del_u\cdot\nabla - \frac{1}{2}(u\cdot\nabla)(\del_u\cdot\del_u) \ . \label{eq:deDonder}
\end{align} 
In terms of this, the general Fronsdal operator \eqref{eq:Fronsdal} becomes:
\begin{align}
 \calF = -\nabla\cdot\nabla + s^2 - 2s - 2 + (g_{\mu\nu}u^\mu u^\nu)(\del_u\cdot\del_u) + (u\cdot\nabla)\calD \ . \label{eq:Fronsdal_deDonder}
\end{align}
The condition $\calD h^{(s)} = 0$ is known as de Donder gauge. While we won't work in this gauge directly, in section \ref{sec:Phi} we will construct a gauge for the Didenko-Vasiliev solution in $EAdS_4$ whose flat-spacetime limit is de Donder.

\subsection{Boundary geometry}

The 3d boundary of $EAdS_4$ is given by the projective lightcone in $\bbR^{1,4}$, i.e. by null vectors $\ell^\mu\in\bbR^{1,4}$, $\ell\cdot\ell = 0$, modulo rescalings $\ell^\mu \cong \rho\ell^\mu$. Boundary quantities will transform under such rescalings as $(\ell\cdot\del_\ell)f = -\Delta f$, according to their conformal weights $\Delta$. We describe 3d vectors at a boundary point $\ell^\mu$ as 5d vectors $\lambda^\mu$ that satisfy $\lambda\cdot\ell = 0$, modulo shifts $\lambda^\mu\cong \lambda^\mu + \alpha\ell^\mu$. For a boundary scalar $f(\ell)$ with weight $\Delta=\frac{1}{2}$, we can define the conformal Laplacian $\Box_\ell f$. In the embedding-space language, this is the same as the 5d d'Alambertian $(\del_\ell\cdot\del_\ell)f$, provided that $f$ is extended away from the $\ell\cdot\ell = 0$ lightcone in a way that preserves the scaling law $(\ell\cdot\del_\ell)f = -\frac{1}{2}f$. The operator $\Box_\ell$ itself has conformal weight 2.

Quantities $f(\ell)$ with weight $\Delta=3$ can be integrated over the boundary as $\int f(\ell) d^3\ell$. With respect to this integration, one can define delta functions $\delta^{\Delta,3-\Delta}(\ell,\ell')$, where the superscripts denote the conformal weights with respect to each of the arguments $(\ell,\ell')$. For a function $f(\ell)$ of weight $\Delta$, or $g(\ell)$ of weight $3-\Delta$, we then have:
\begin{align}
 \int d^3\ell' \delta^{\Delta,3-\Delta}(\ell,\ell')f(\ell') = f(\ell) \ ; \quad  \int d^3\ell\,\delta^{\Delta,3-\Delta}(\ell,\ell')g(\ell) = g(\ell') \ .
\end{align}
Choosing a \emph{conformal frame} on the boundary is equivalent to choosing a \emph{section} of the $\bbR^{1,4}$ lightcone. The simplest such section is a flat one, parameterized by Poincare coordinates $\mathbf{y}$:
\begin{align}
 \ell^\mu(\mathbf{y}) = \left(\frac{1+\mathbf{y}^2}{2}, \frac{1-\mathbf{y}^2}{2}, \mathbf{y} \right) \ ; \quad d\ell\cdot d\ell = \mathbf{dy}^2 \ . \label{eq:flat_section}
\end{align}
In this frame, the distance between two boundary points is:
\begin{align}
 \sqrt{(\ell-\ell')\cdot(\ell-\ell')} = \sqrt{-2\ell\cdot\ell'} = |\mathbf{y-y'}| \ , \label{eq:flat_distance}
\end{align}
and the conformal Laplacian $\Box_\ell$ is just the flat Laplacian:
\begin{align}
 \Box_\ell = \del_{\mathbf{y}}\cdot\del_{\mathbf{y}} \ . \label{eq:flat_Laplacian}
\end{align}
Our formulas below will be covariant (i.e. independent of frame choice) as much as possible, but they're occasionally easier to derive in the Poincare frame \eqref{eq:flat_section}. In particular, from \eqref{eq:flat_distance}-\eqref{eq:flat_Laplacian}, we immediately see the important identity:
\begin{align}
 \Box_\ell\frac{1}{\sqrt{-2\ell\cdot\ell'}} = -4\pi\delta^{\frac{5}{2},\frac{1}{2}}(\ell,\ell') \ . \label{eq:Box_1_r}
\end{align}
Poincare coordinates can also be used to parameterize bulk points $x\in EAdS_4$, as:
\begin{align}
 x^\mu(z,\mathbf{y}) = \frac{1}{z}\left(\frac{1+z^2+\mathbf{y}^2}{2}, \frac{1-z^2-\mathbf{y}^2}{2}, \mathbf{y} \right) \ ; \quad dx\cdot dx = \frac{dz^2 + \mathbf{dy}^2}{z^2} \ . \label{eq:Poincare}
\end{align}
The boundary in these coordinates is at $z\rightarrow 0$, where $x^\mu(z,\mathbf{y})$ asymptotes to $\ell^\mu(\mathbf{y})/z$.

\subsection{Boundary CFT and boundary-bulk propagator}

The CFT that lives on our 3d boundary is the free $U(N)$ vector model, defined by the action:
\begin{align}
 S_{\text{CFT}} = -\int d^3\ell\,\bar\chi_I(\ell)\Box_\ell\chi^I(\ell) \ , \label{eq:CFT}
\end{align}
where $I=1\dots N$ is a color index, and $\chi^I(\ell)$ with its complex conjugate $\bar\chi_I(\ell)$ are scalar fields with conformal weight $\Delta = \frac{1}{2}$. The propagator for these fundamental fields can be read off from \eqref{eq:Box_1_r} as:
\begin{align}
 G_{\text{CFT}}(\ell,\ell') = \frac{1}{4\pi\sqrt{-2\ell\cdot\ell'}} \ ; \quad \Box_\ell G_{\text{CFT}}(\ell,\ell') = -\delta^{\frac{5}{2},\frac{1}{2}}(\ell,\ell') \ . \label{eq:G_CFT}
\end{align}
The fundamental single-trace operators in the theory \eqref{eq:CFT} are the \emph{bilocals}:
\begin{align}
 \calO(\ell,\ell') \equiv \frac{\chi^I(\ell)\bar\chi_I(\ell')}{\sqrt{N}} \ . \label{eq:bilocal}
\end{align}
By Taylor-expanding these around $\ell=\ell'$, we obtain the local single-trace primaries, i.e. the tower of HS currents \cite{Craigie:1983fb,Anselmi:1999bb,David:2020ptn} (including the honorary spin-0 ``current'' $\bar\chi_I(\ell)\chi^I(\ell)/\sqrt{N}$). These local currents can be encoded conveniently by contracting their indices with a null polarization vector $\lambda^\mu$ at $\ell^\mu$, satisfying $\lambda\cdot\lambda = \lambda\cdot\ell = 0$:
\begin{align}
  j^{(s)}(\ell,\lambda) = \lambda^{\mu_1}\dots\lambda^{\mu_s} j_{\mu_1\dots\mu_s}(\ell) \ .
\end{align}
The currents' relation to the bilocal \eqref{eq:bilocal} is then expressed compactly via a differential operator $D^{(s)}$, as:
\begin{align}
 j^{(s)}(\ell,\lambda) &= \left.D^{(s)}(\del_\ell,\del_{\ell'},\lambda)\,\calO(\ell,\ell')\right|_{\ell=\ell'} \ ; \label{eq:j} \\
 D^{(s)}(\del_\ell,\del_{\ell'},\lambda) &= i^s\sum_{m=0}^s (-1)^m \binom{2s}{2m} (\lambda\cdot\del_\ell)^m (\lambda\cdot\del_{\ell'})^{s-m} \ . \label{eq:D}
\end{align}
The connected correlators of bilocals \eqref{eq:bilocal} are given by simple 1-loop Feynman diagrams composed of propagators \eqref{eq:G_CFT}:
\begin{align}
 \left<\calO(\ell_1,\ell_1')\dots \calO(\ell_n,\ell_n')\right>_{\text{connected}} = N^{\frac{2-n}{2}}\left(\prod_{p=1}^n G(\ell_p',\ell_{p+1}) + \text{permutations}\right) \ , \label{eq:correlators_bilocal}
\end{align}
where the product is cyclic, i.e. $\ell_{n+1}\equiv\ell_1$ and the sum is over cyclically inequivalent permutations of $(1,\dots,n)$. From these, one can derive the correlators of local currents, via the Taylor expansion \eqref{eq:j}.

The boundary-bulk propagators dual to the boundary HS currents \eqref{eq:j} read \cite{Mikhailov:2002bp}:
\begin{align}
 \Pi^{(s)}(x,u;\ell,\lambda) = -\frac{C_s}{s!}\,\frac{\big[(\lambda\cdot x)(\ell\cdot u) - (\ell\cdot x)(\lambda\cdot u)\big]^s}{(\ell\cdot x)^{2s+1}} \ , \label{eq:Pi}
\end{align}
with the normalization factors first given correctly in \cite{Costa:2014kfa} (see also \cite{Halpern:2015zia,David:2020fea}):
\begin{align}
 C_s = \frac{1}{4\pi^2}\times \left\{
  \begin{array}{cl}
    1 & \qquad s=0 \\[0.3em]
    \displaystyle \frac{2^{s+1}(s!)^2}{(2s)!} & \qquad s\geq 1 
  \end{array} \right. \ .
\end{align}
With respect to its bulk arguments $(x,u)$, the propagator $\Pi^{(s)}$ satisfies the standard constraints \eqref{eq:h_tensor_properties} for a Fronsdal field, as well as the traceless and transverse gauge conditions $(\del_u\cdot\del_u)\Pi^{(s)} = (\del_u\cdot\nabla)\Pi^{(s)} = 0$. With respect to its boundary arguments $(\ell,\lambda)$, $\Pi^{(s)}$ has the same conformal weight $(\ell\cdot\del_\ell)\Pi^{(s)} = -(s+1)\Pi^{(s)}$ and tensor rank $(\lambda\cdot\del_\lambda)\Pi^{(s)} = s\Pi^{(s)}$ as the boundary currents \eqref{eq:j}, and is invariant under the gauge-like shift symmetry $\lambda^\mu\rightarrow \lambda^\mu + \alpha\ell^\mu$.

\subsection{Bulk geodesics}

The Didenko-Vasiliev solution is the field of an HS-charged source concentrated on a bulk geodesic. Before describing the solution and its properties, it will be useful to discuss bulk geodesics in their own right.

\subsubsection{Geodesic geometry and related formulas}

A geodesic in $EAdS_4$ is a hyperbola in the $\bbR^{1,4}$ embedding space. The hyperbola's asymptotes are two lightrays through the origin in $\bbR^{1,4}$, or, equivalently, two points on the conformal boundary of $EAdS_4$. In fact, (oriented) bulk geodesics are in one-to-one correspondence with (ordered) pairs of boundary points. We can parameterize a geodesic's boundary endpoints by two lightlike vectors $\ell^\mu,\ell'^\mu$, keeping in mind the usual redundancy of such vectors under rescalings. The geodesic itself can then be parameterized as:
\begin{align}
 \gamma(\ell,\ell'):\quad x^\mu(\tau;\ell,\ell') = \frac{e^\tau \ell^\mu + e^{-\tau}\ell'^\mu}{\sqrt{-2\ell\cdot\ell'}} \ , \label{eq:geodesic}
\end{align}
with $\tau$ a proper-length parameter. The geodesic \eqref{eq:geodesic} is the intersection in $\bbR^{1,4}$ of the $EAdS_4$ hyperboloid with the 2d plane spanned by $\ell^\mu,\ell'^\mu$.

The distance of a bulk point $x\in EAdS_4$ from a geodesic $\gamma(\ell,\ell')$ can be parameterized by the function:
\begin{align}
 R(x;\ell,\ell') = \sqrt{-\frac{2(\ell\cdot x)(\ell'\cdot x)}{\ell\cdot\ell'} - 1} \ , \label{eq:R}
\end{align}
which is invariant under rescalings of $\ell^\mu,\ell'^\mu$. This $R(x;\ell,\ell')$ is just the flat $\bbR^{1,4}$ distance between the point $x^\mu$ and the $(\ell,\ell')$ plane. It is related to the geodesic $EAdS_4$ distance $\chi$ as $R = \sinh\chi$. 

We can define a delta function that localizes $x\in EAdS_4$ on the geodesic $\gamma(\ell,\ell')$, i.e. at $R=0$, as:
\begin{align}
 \delta^3(x;\ell,\ell') = \int_{-\infty}^\infty d\tau\,\delta^4(x,x(\tau;\ell,\ell')) \ , \label{eq:geodesic_delta}
\end{align}
where $\delta^4$ is the delta function on $EAdS_4$, and $x(\tau;\ell,\ell')$ is the proper-length parameterization \eqref{eq:geodesic} of $\gamma(\ell,\ell')$. The geodesic delta function $\delta^3(x;\ell,\ell')$ is invariant under rescalings of $\ell^\mu,\ell'^\mu$. 

The distance function \eqref{eq:R} and the geodesic delta function $\delta^3(x;\ell,\ell')$ are related by the Laplacian identity:
\begin{align}
 (\nabla\cdot\nabla + 2)\frac{1}{R} = -4\pi\delta^3(x;\ell,\ell') \ .
\end{align}
There is a useful alternative formula for the distance function \eqref{eq:R}:
\begin{align}
 R(x;\ell,\ell') = \sqrt{\frac{\ell_\mu X^{\mu\nu}\ell'_\nu}{\ell\cdot\ell'}} \ ,
\end{align}
where $X^\mu_\nu$ is the reflection matrix with respect to $x^\mu$ in the $\bbR^{1,4}$ embedding space:
\begin{align}
X_\mu^\nu(x) = -\delta_\mu^\nu - 2x_\mu x^\nu \ .
\end{align}
In particular, the following conditions are equivalent:
\begin{align}
 R(x;\ell,\ell') = 0 \quad \Longleftrightarrow \quad x\text{ lies on }\gamma(\ell,\ell') \quad \Longleftrightarrow \quad X^\mu_\nu(x)\ell^\nu \sim \ell'^\mu \ . \label{eq:coincidence}
\end{align}
This means that the geodesic delta function \eqref{eq:geodesic_delta} should be proportional to a \emph{boundary} delta function with support on $\ell^\mu\sim X^\mu_\nu(x)\ell'^\nu$. Specifically, for any weight $\Delta$, we have:
\begin{align}
 \delta^{\Delta,3-\Delta}(\ell,X\ell') = \frac{\delta^3(x;\ell,\ell')}{(-2\ell\cdot x)^\Delta(-2\ell'\cdot x)^{3-\Delta}} \ , \label{eq:delta_proportionality}
\end{align}
where the coefficient can be recast into other forms by employing the constraint $R(x;\ell,\ell') = 0$, i.e. $2(\ell\cdot x)(\ell'\cdot x) = -\ell\cdot\ell'$. In the special case $\Delta=\frac{1}{2}$, when the Laplacian $\Box_\ell\delta^{\Delta,3-\Delta}(\ell,X\ell')$ is well-defined, we can also relate the bulk and boundary Laplacians as:
\begin{align}
 \Box_\ell\delta^{\frac{1}{2},\frac{5}{2}}(\ell,X\ell') = \frac{(\nabla\cdot\nabla)\delta^3(x;\ell,\ell')}{(-2\ell\cdot\ell')^{5/2}} \ . \label{eq:Box_delta_proportionality}
\end{align}

\subsubsection{Derivation of delta-function relationships}

Let us derive eqs. \eqref{eq:delta_proportionality}-\eqref{eq:Box_delta_proportionality}. We already argued that \eqref{eq:delta_proportionality} must be true up to a proportionality coefficient. The coefficient is then fixed up to a numerical factor by the weights with respect to $\ell^\mu$ and $\ell'^\mu$. To fix this remaining factor, it is easiest to use Poincare coordinates \eqref{eq:flat_section},\eqref{eq:Poincare}. Any triple of points $x,\ell,\ell'$ can be expressed, up to $\bbR^{1,4}$ rotations and rescalings of $\ell^\mu$, as:
\begin{gather}
 x^\mu = x^\mu(z,\mathbf{y}) = \frac{1}{z}\left(\frac{1+z^2+\mathbf{y}^2}{2}, \frac{1-z^2-\mathbf{y}^2}{2}, \mathbf{y} \right) \ ; \label{eq:varying_x_ell_1} \\ 
 \ell^\mu = \ell^\mu(\mathbf{p}) = \left(\frac{1 + \mathbf{p}^2}{2}, \frac{1-\mathbf{p}^2}{2}, \mathbf{p} \right) \ ; \quad \ell'^\mu = \left(\frac{1}{2}, -\frac{1}{2}, \vec{0} \right) \ , \label{eq:varying_x_ell_2}
\end{gather}
where $\ell'^\mu$ is fixed to the ``point at infinity'' of the flat boundary frame \eqref{eq:flat_section}. The three points lie on the same geodesic when $\mathbf{y} = \mathbf{p}$, which also corresponds to $\ell^\mu = z^2X^\mu{}_\nu\ell'^\nu$. The geodesic itself -- the line of varying $z$ at constant $\mathbf{y}$ -- is perpendicular to the 3d bulk hypersurfaces of varying $\mathbf{y}$ at fixed $z$. The boundary measure $d^3\ell$ is just the flat measure $d^3\mathbf{p}$, whereas the bulk measure on a constant-$z$ hypersurface is $d^3\mathbf{y}/z^3$. Thus, in the parameterization \eqref{eq:varying_x_ell_1}-\eqref{eq:varying_x_ell_2}, the two delta functions in \eqref{eq:delta_proportionality} become:
\begin{align}
 \delta^{\Delta,3-\Delta}(\ell,X\ell') &= z^{2(3-\Delta)}\delta^{\Delta,3-\Delta}(\ell,z^2X\ell') = z^{2(3-\Delta)}\delta^3(\mathbf{y}-\mathbf{p}) \ ; \\
 \delta^3(x;\ell,\ell') &= z^3\delta^3(\mathbf{y}-\mathbf{p}) \ .
\end{align}
This is indeed consistent with the coefficient written in \eqref{eq:delta_proportionality}, since at $\mathbf{y} = \mathbf{p}$ in the parameterization \eqref{eq:varying_x_ell_1}-\eqref{eq:varying_x_ell_2}, we have: 
\begin{align}
 \ell\cdot\ell' = -\frac{1}{2} \ ; \quad \ell\cdot x = -\frac{z}{2} \ ; \quad \ell'\cdot x = -\frac{1}{2z} \ . \label{eq:flat_scalar_products}
\end{align}
Now, let us turn to the special case $\Delta = \frac{1}{2}$ and the Laplacian formula \eqref{eq:Box_delta_proportionality}. In the parameterization \eqref{eq:varying_x_ell_1}-\eqref{eq:varying_x_ell_2}, $\Box_\ell$ becomes just the flat 3d Laplacian $\del_\mathbf{p}\cdot\del_\mathbf{p}$, and we get:
\begin{align}
 \Box_\ell\delta^{\frac{1}{2},\frac{5}{2}}(\ell,X\ell') = z^5(\del_\mathbf{p}\cdot\del_\mathbf{p})\delta^3(\mathbf{y}-\mathbf{p}) \ . \label{eq:boundary_Box_delta}
\end{align}
On the bulk side, we write out the $EAdS_4$ Laplacian in Poincare coordinates, getting:
\begin{align}
 \begin{split}
   (\nabla\cdot\nabla)\delta^3(x;\ell,\ell') &= z^4\left(\del_z\!\left(\frac{1}{z^2}\del_z\delta^3(x;\ell,\ell') \right) + \frac{1}{z^2}(\del_\mathbf{y}\cdot\del_\mathbf{y})\delta^3(x;\ell,\ell') \right) \\
     &= 0 + z^5(\del_\mathbf{y}\cdot\del_\mathbf{y})\delta^3(\mathbf{y}-\mathbf{p}) \ .
 \end{split}
\end{align}
Comparing with \eqref{eq:boundary_Box_delta} and re-inserting the scalar products \eqref{eq:flat_scalar_products} to recover the correct conformal weights, we arrive at \eqref{eq:Box_delta_proportionality}.

\subsection{Linearized Didenko-Vasiliev solution}

The Didenko-Vasiliev solution \cite{Didenko:2009td} is a solution of the non-linear Vasiliev equations, structurally similar to supergravity's BPS black holes. We will be interested here in the solution's linearized version \cite{Didenko:2008va}, which consists of a multiplet of Fronsdal fields (one for each spin), satisfying the Fronsdal field equation \eqref{eq:Fronsdal_equation} with a \emph{particle-like source} concentrated on a bulk geodesic $\gamma(\ell,\ell')$. 

To write the solution explicitly, we will use some geometric structures that relate a bulk point $x$ to a geodesic $\gamma(\ell,\ell')$. These include the distance function $R(x;\ell,\ell')$ from \eqref{eq:R}, along with a pair of $EAdS_4$ vectors at $x$:
\begin{align}
 t_\mu(x;\ell,\ell') &= \frac{1}{2}\left(\frac{\ell'_\mu}{\ell'\cdot x} - \frac{\ell_\mu}{\ell\cdot x} \right) \ ; \label{eq:t} \\ 
 r_\mu(x;\ell,\ell') &= x_\mu + \frac{1}{2}\left(\frac{\ell_\mu}{\ell\cdot x} + \frac{\ell'_\mu}{\ell'\cdot x} \right) \ ; \label{eq:r}
\end{align}
Here, $r^\mu(x;\ell,\ell')$ points radially away from the $\gamma(\ell,\ell')$ geodesic, while $t^\mu(x;\ell,\ell')$ points ``parallel to'' $\gamma(\ell,\ell')$, in the sense of parallel transport along $r^\mu$. These vectors satisfy:
\begin{align}
 t\cdot x = r\cdot x = t\cdot r = 0 \ ; \quad t\cdot t = \frac{1}{1+R^2} \ ; \quad r\cdot r = \frac{R^2}{1+R^2} \ .
\end{align}
We then construct a \emph{complex null} vector in the $(t,r)$ plane:
\begin{align}
 k_\mu(x;\ell,\ell') &= \frac{1}{2}\left(t_\mu + \frac{ir_\mu}{R} \right) \ ; \quad k\cdot k = 0 \ ; \quad (k\cdot\nabla)k_\mu = 0 \ . \label{eq:k}
\end{align}
In Lorentzian signature, $k^\mu$ would be a real, affine tangent to radial lightrays emanating from $\gamma(\ell,\ell')$. The $EAdS_4$ derivatives of $R,t^\mu,r^\mu,k^\mu$ (at $R\neq 0$) are listed in eqs. (70)-(71) of \cite{David:2020fea}; we will recall the necessary formulas as we go along.

With these building blocks in hand, the Didenko-Vasiliev solution is now given by the following multiplet of Fronsdal fields:
\begin{align}
  \phi^{(s)}(x,u;\ell,\ell') = \frac{1}{8\pi^2 R \sqrt{-\ell\cdot\ell'}} \times \left\{
    \begin{array}{cl}
      1 & \qquad s = 0 \\
      \displaystyle \frac{2}{s!}(i\sqrt{2})^s(u\cdot k)^s & \qquad s\geq 1
    \end{array} \right. \ . \label{eq:phi}
\end{align}
$\phi^{(s)}$ has weight $\Delta=\frac{1}{2}$ in each of its boundary arguments $(\ell,\ell')$. In its bulk arguments $(x,u)$, $\phi^{(s)}$ satisfies the standard constraints \eqref{eq:h_tensor_properties} for a Fronsdal field, as well as the traceless gauge condition $(\del_u\cdot\del_u)\phi^{(s)} = 0$. Though the potentials \eqref{eq:phi} are complex, their gauge-invariant curvatures are always real (for even $s$) or imaginary (for odd $s$). In other words, for even/odd $s$, the imaginary/real part of \eqref{eq:phi} respectively is pure-gauge. The solution's particular form \eqref{eq:phi} (up to an overall, $s$-independent normalization) has been worked out in \cite{David:2020fea}. In particular, it was shown there how the master-field structure from \cite{Didenko:2009td}, when translated into canonically normalized Fronsdal fields, implies the particular $s$-dependent normalizations of \eqref{eq:phi}. In the absence of an action in Vasiliev's master-field language, this normalization-fixing was mediated by the 2-point functions of local boundary currents, which can be expressed through HS star products of master fields as in \cite{Neiman:2017mel}, building on the methods of \cite{Colombo:2012jx,Didenko:2012tv}.

The Einstein curvature \eqref{eq:Einstein} of $\phi^{(s)}$, i.e. the bulk source in its Fronsdal equation \eqref{eq:Fronsdal_equation}, reads:
\begin{align}
 \calG \phi^{(s)}(x;\ell,\ell') = \frac{(i\sqrt{2})^s}{2\pi s!\sqrt{-\ell\cdot\ell'}}\,\delta^3(x;\ell,\ell') \big[(u\cdot t)^s - \text{double traces} \big] \ . \label{eq:Einstein_phi}
\end{align}
Here, $\delta^3(x;\ell,\ell')$ is the geodesic delta function \eqref{eq:geodesic_delta}, with support on $R=0$; $t^\mu$ is the vector \eqref{eq:t}, which at $R=0$ becomes just the unit tangent to $\gamma(\ell,\ell')$; and ``${}-\text{double traces}$'' means that we subtract $\sim (g_{\mu\nu}u^\mu u^\nu)^2$ pieces so as to satisfy the double-tracelessness condition $(\del_u\cdot\del_u)^2\calG \phi^{(s)} = 0$. Eq. \eqref{eq:Einstein_phi} shows explicitly the HS charges carried by the geodesic. The $s$-dependent coefficients in \eqref{eq:Einstein_phi} encode the BPS-like proportionality pattern between the charges of different spins. In terms of the traceless structure \eqref{eq:T}, the Einstein tensor \eqref{eq:Einstein_phi} and the corresponding Fronsdal tensor can be written as:
\begin{align}
  \calG \phi^{(s)} &= \frac{(i\sqrt{2})^s\,\delta^3(x;\ell,\ell')}{2\pi\sqrt{-\ell\cdot\ell'}} \left(\calT^{(s)}(x,t,u) + \frac{\theta(s-2)}{4s}(g_{\mu\nu}u^\mu u^\nu)\calT^{(s-2)}(x,t,u) \right) \ ; \\
  \calF \phi^{(s)} &= \frac{(i\sqrt{2})^s\,\delta^3(x;\ell,\ell')}{2\pi\sqrt{-\ell\cdot\ell'}} \left(\calT^{(s)}(x,t,u) - \frac{\theta(s-2)}{4s(s-1)}(g_{\mu\nu}u^\mu u^\nu)\calT^{(s-2)}(x,t,u) \right) \ , \label{eq:Fronsdal_phi}
\end{align}
where $\theta$ is the step function:
\begin{align}
 \theta(p) = \left\{
   \begin{array}{cl}
     1 & \qquad p\geq 0 \\
     0 & \qquad p<0
   \end{array} \right. \ ,
\end{align}
and we assume the convention that $\theta(p)$ for negative $p$ vanishes ``stronger than anything else'', so that e.g. $\frac{\theta(s-2)}{s(s-1)}$ is zero for $s=0,1$.

\subsection{Mutual action of two Didenko-Vasiliev solutions} \label{sec:preliminaries:DV_mutual}

It was recently understood \cite{David:2020fea} that the Didenko-Vasiliev multiplet \eqref{eq:phi} forms the \emph{bulk dual of the boundary bilocal operator} \eqref{eq:bilocal}. This identification was motivated by the solution's Penrose transform, and its behavior within HS algebra -- concepts that we won't introduce in this paper. However, the main result of \cite{David:2020fea} is formulated in terms of Fronsdal fields. It concerns the mutual bulk action \eqref{eq:on_shell_action} between a pair of Didenko-Vasiliev solutions $\phi^{(s)}(x;\ell,\ell),\phi^{(s)}(x;L,L')$, i.e. the action \eqref{eq:on_shell_action} evaluated for $h^{(s)}(x,u) = \phi^{(s)}(x,u;\ell,\ell')+\phi^{(s)}(x,u;L,L')$ with self-interaction neglected, i.e. keeping only the $\sim\phi^{(s)}(\ell,\ell')\phi^{(s)}(L,L')$ cross-terms:
\begin{align}
 S = -\frac{1}{2}\int_{EAdS_4} d^4x \sum_{s=0}^\infty s!\,\phi^{(s)}(x,\del_u;\ell,\ell')\,\calG\phi^{(s)}(x,u;L,L') + \Big((\ell,\ell')\leftrightarrow(L,L')\Big) \ .
\end{align}
By spacetime symmetry, the $(\ell,\ell')\leftrightarrow(L,L')$ term must be equal to the original one, so we can just discard it at the cost of removing the factor of $\frac{1}{2}$:
\begin{align}
 S = -\int_{EAdS_4} d^4x \sum_{s=0}^\infty s!\,\phi^{(s)}(x,\del_u;\ell,\ell')\,\calG\phi^{(s)}(x,u;L,L') \ . \label{eq:mutual_S}
\end{align}
Now, the statement of \cite{David:2020fea} is that the bulk action \eqref{eq:mutual_S} equals minus the connected correlator between the corresponding boundary bilocals $\calO(\ell,\ell'),\calO(L,L')$, i.e.:
\begin{align}
 \int_{EAdS_4} d^4x \sum_{s=0}^\infty s!\,\phi^{(s)}(x,\del_u;\ell,\ell')\,\calG\phi^{(s)}(x,u;L,L') = \left<\calO(\ell,\ell')\calO(L,L')\right>_{\text{connected}} \ . \label{eq:bilocal_2_point_raw}
\end{align}
Evaluating the LHS via \eqref{eq:Einstein_phi} and the RHS via \eqref{eq:correlators_bilocal} (and multiplying both sides by $2\pi\sqrt{-L\cdot L'}$), the statement becomes:
\begin{align}
 \int_{-\infty}^\infty d\tau \sum_{s=0}^\infty s!(i\sqrt{2})^s \phi^{(s)}\big(x(\tau;L,L'),\dot x(\tau;L,L');\ell,\ell'\big) = \frac{1}{16\pi}\sqrt{\frac{-L\cdot L'}{(\ell\cdot L')(L\cdot\ell')}} \ . \label{eq:bilocal_2_point}
\end{align}
Here, $x(\tau;L,L')$ is the proper-length parameterization \eqref{eq:geodesic} of the $\gamma(L,L')$ geodesic, while $\dot x^\mu(\tau;L,L')$ is the unit tangent to it:
\begin{align}
 \dot x^\mu(\tau;L,L') \equiv \frac{d}{d\tau} x^\mu(\tau;L,L') = \frac{e^\tau L^\mu - e^{-\tau}L'^\mu}{\sqrt{-2L\cdot L'}} = t^\mu\big(x^\mu(\tau;L,L');L,L'\big) \ . \label{eq:x_dot}
\end{align}
It should be possible to derive the result \eqref{eq:bilocal_2_point} as a special case of the theory of (spinning) geodesic Witten diagrams \cite{Hijano:2015zsa,Dyer:2017zef}. In \cite{David:2020fea}, we simply checked it for various choices of $(\ell,\ell',L,L')$, by performing the sum over spins in \eqref{eq:bilocal_2_point} analytically, and then the integral numerically. The original version of \cite{David:2020fea} claimed erroneously that eq. \eqref{eq:bilocal_2_point} is only true after we remove the pure-gauge imaginary/real part of $\phi^{(s)}$, for even/odd $s$ respectively. In fact, this imaginary/real part is harmless, as its contribution cancels between the two halves of the integral \eqref{eq:bilocal_2_point}, on either side of the point of closest approach between $\gamma(\ell,\ell')$ and $\gamma(L,L')$.

In section \ref{sec:analytic_2_point}, we will present a simple analytic derivation of \eqref{eq:bilocal_2_point}, employing a new, non-traceless gauge for the Didenko-Vasiliev solution.

\section{Boundary-local limit} \label{sec:local_limit}

\subsection{Limiting behavior of Didenko-Vasiliev potential} \label{sec:local_limit:potential}

If the Didenko-Vasiliev solution \eqref{eq:phi} is indeed the bulk dual of the boundary bilocal \eqref{eq:bilocal}, then in the $\ell=\ell'$ limit \eqref{eq:j}-\eqref{eq:D}, where the boundary bilocal reduces to local currents, the DV solution should reduce to the familiar boundary-bulk propagators. In \cite{David:2020ptn}, this was shown to be the case for the solution's \emph{Weyl curvature}. Here, we will show that the same is true for the potentials \eqref{eq:phi} themselves. Let us expand the formula \eqref{eq:phi} as:
\begin{align}
 \phi^{(s)}(x,u;\ell,\ell') &= \frac{1}{8\pi^2 R \sqrt{-\ell\cdot\ell'}} \times \left\{
   \begin{array}{cl}
     1 & \qquad s = 0 \\[0.3em]
     \displaystyle \frac{2}{s!} \left(\frac{i}{\sqrt{2}}\right)^s \left(u\cdot t + i\,\frac{u\cdot r}{R}\right)^s & \qquad s\geq 1
   \end{array} \right. \ . \label{eq:phi_t_r}
\end{align}
We now act on these fields with the differential operator $D^{(\tilde s)}(\del_\ell,\del_{\ell'},\lambda)$ from \eqref{eq:D}, and then set $\ell=\ell'$. We make the following observations:
\begin{itemize}
	\item In the $\ell=\ell'$ limit, $\sqrt{-\ell\cdot\ell'}$ and $t^\mu$ vanish, $r^\mu$ stays finite, and $R$ diverges. The product $R \sqrt{-\ell\cdot\ell'}$ tends to the finite value $-\sqrt{2}(\ell\cdot x)$.
	\item Due to the constraints $\lambda\cdot\lambda = \lambda\cdot\ell = 0$, all contributions from $\ell,\ell'$ derivatives acting on the $\frac{1}{\sqrt{-\ell\cdot\ell'}}$ factor vanish.
	\item All contributions from $\ell,\ell'$ derivatives acting on $\frac{1}{R}$ factors vanish, since they result in higher powers of $\frac{1}{R}$.
\end{itemize}
The upshot is that $\left.D^{(\tilde s)}\phi^{(s)}\right|_{\ell=\ell'}$ is non-vanishing only for $s=\tilde s$, and then all the $\ell,\ell'$ derivatives must be acting on the factors of $t^\mu$. The result is then easy to evaluate, giving:
\begin{align}
 \begin{split}
   [D^{(\tilde s)}\phi^{(s)}](x,u;\ell,\lambda) &= -\delta_{s,\tilde s}\,\frac{(\sqrt{2})^{s-1}}{8\pi^2} \frac{\big[(\lambda\cdot x)(\ell\cdot u) - (\ell\cdot x)(\lambda\cdot u)\big]^s}{(\ell\cdot x)^{2s+1}} \\
     &= \delta_{s,\tilde s}\,\frac{(\sqrt{2})^{s-1}s!}{8\pi^2 C_s}\,\Pi^{(s)}(x,u;\ell,\lambda) \ ,
 \end{split} \label{eq:local_limit}
\end{align}
where $\delta_{s,\tilde s}$ is a Kronecker symbol imposing $s=\tilde s$, and $\Pi^{(s)}$ is the boundary-bulk propagator \eqref{eq:Pi}. Thus, the boundary-local limit $\left.D^{(\tilde s)}\phi^{(s)}\right|_{\ell=\ell'}$ indeed reproduces the boundary-bulk propagators, up to normalization.

\subsection{Limiting behavior of mutual action} \label{sec:local_limit:action}

In this section, we prove a limiting case of the formula \eqref{eq:bilocal_2_point_raw}-\eqref{eq:bilocal_2_point} for the mutual action of two Didenko-Vasiliev solutions. This will serve as a component in the full proof, in section \ref{sec:analytic_2_point}. 

Let us apply the boundary-local limit \eqref{eq:local_limit} to $\phi^{(s)}(x;\ell,\ell')$ in eq. \eqref{eq:bilocal_2_point}. In other words, let us act on \eqref{eq:bilocal_2_point} with the differential operator $D^{(\tilde s)}(\del_\ell,\del_\ell',\lambda)$, and set $\ell=\ell'$. Rearranging some numerical factors onto the RHS, the resulting limit of \eqref{eq:bilocal_2_point} reads:
\begin{align}
 \begin{split}
   &\frac{s!}{C_s}\int_{-\infty}^\infty d\tau\,\Pi^{(s)}\big(x(\tau;L,L'),\dot x(\tau;L,L');\ell,\lambda\big) \\
   &\quad = \frac{\pi(2s)!}{8^s\sqrt{2}(s!)^2}\big[(\ell\cdot L)(\lambda\cdot L') - (\ell\cdot L')(\lambda\cdot L)\big]^s\sqrt{\frac{-L\cdot L'}{(\ell\cdot L)^{2s+1}(\ell\cdot L')^{2s+1}}} \ .
 \end{split} \label{eq:bilocal_local_2_point}
\end{align}
Up to a numerical factor, the structure of eq. \eqref{eq:bilocal_local_2_point} is completely fixed by the scaling weight $\Delta=s+1$ with respect to $\ell^\mu$, the correct power $s$ of $\lambda^\mu$, the shift symmetry $\lambda^\mu\rightarrow\lambda^\mu+\alpha\ell^\mu$, and invariance under rescalings of $L^\mu,L'^\mu$. Thus, it remains to establish the numerical coefficient. To do this, let us fix the parameter vectors $(\ell,\lambda,L,L')$ such that:
\begin{align}
 L\cdot L' = \ell\cdot L = \ell\cdot L' = -\frac{1}{2} \ ; \quad \lambda\cdot L = \frac{1}{2} \ ; \quad \lambda\cdot L' = -\frac{1}{2} \ . \label{eq:ell_L_lambda_products}
\end{align}
We then get, for the position vector $x^\mu(\tau;L,L')$ and unit tangent $\dot x^\mu(\tau;L,L')$ on $\gamma(L,L')$:
\begin{align}
 \ell\cdot x = -\cosh\tau \ ; \quad \ell\cdot\dot x = -\sinh\tau \ ; \quad \lambda\cdot x = \sinh\tau \ ; \quad \lambda\cdot\dot x = \cosh\tau \ .
\end{align}
With these simplifications, the integral in \eqref{eq:bilocal_local_2_point} can be evaluated as:
\begin{align}
 &\frac{s!}{C_s}\int_{-\infty}^\infty d\tau\,\Pi^{(s)}\big(x(\tau;L,L'),\dot x(\tau;L,L');\ell,\lambda\big) = -\int_{-\infty}^\infty d\tau\,\frac{\big[(\lambda\cdot x)(\ell\cdot\dot x) - (\ell\cdot x)(\lambda\cdot\dot x)\big]^s}{(\ell\cdot x)^{2s+1}} \nonumber \\
 &\quad = \int_{-\infty}^\infty \frac{d\tau}{\cosh^{2s+1}\tau} = \int_{-\pi/2}^{\pi/2} d\beta \cos^{2s}\beta = \frac{\pi(2s)!}{4^s(s!)^2} \ ,
\end{align}
where we changed the integration variable as $\frac{1}{\cosh\tau}\equiv\cos\beta$. Recalling \eqref{eq:ell_L_lambda_products}, we see that this indeed agrees with the numerical coefficient in \eqref{eq:bilocal_local_2_point}. 

An alternative proof of \eqref{eq:bilocal_local_2_point} would be to recast the bulk action not as an integral of $\Pi^{(s)}(x,u;\ell,\lambda)$ over $\gamma(L,L')$, but an evaluation of the boundary data of $\phi^{(s)}(x,u;L,L')$ at $\ell$.

\section{Boundary Laplacian} \label{sec:Box}

\subsection{Results}

The boundary bilocal \eqref{eq:bilocal} inherits the free field equation $\Box_\ell\chi^I(\ell) = \Box_{\ell'}\bar\chi_I(\ell') = 0$ from the fundamental field on each of its ``legs''. Inside correlators, this means that e.g. $\Box_\ell\calO(\ell,\ell')$ vanishes \emph{up to contact terms}. If the Didenko-Vasiliev solution \eqref{eq:phi} is the bulk dual of $\calO(\ell,\ell')$, then it must also satisfy such an equation. Indeed, in \cite{Neiman:2017mel} we found a $\Box_\ell = 0$ relation for the Weyl curvature of \eqref{eq:phi}, \emph{away from the $R=0$ singularity}. Here, we'll derive the corresponding result for the potentials \eqref{eq:phi} themselves, including the detailed behavior (in terms of delta functions) at $R=0$.

First, let us state the explicit result for $\Box_\ell\phi^{(s)}$, as a delta function (for $s=0$), or the sum of a delta function and an ordinary rational function (for $s\geq 1$):
\begin{align}
  &s=0: \quad \Box_\ell \phi^{(0)} = -\frac{\delta^3(x;\ell,\ell')}{8\pi(\ell\cdot x)^2\sqrt{-\ell\cdot\ell'}} \ ; \label{eq:Box_phi_0} \\
  &s\geq 1: \quad \Box_\ell \phi^{(s)} = \frac{2(i\sqrt{2})^s}{s+1}\,\calT^{(s)}(x,t,u)\,\Box_\ell \phi^{(0)} - \frac{(i\sqrt{2})^s(1+iR)^2}{16\pi^2(s-1)! R^3(\ell\cdot x)^2\sqrt{-\ell\cdot\ell'}} \label{eq:Box_phi_s} \\
   &\qquad \times \left((s-1+2iR)(u\cdot k)^s - \frac{s-1}{4}(g_{\mu\nu}u^\mu u^\nu)(u\cdot k)^{s-2} - \frac{is(1+R^2)}{R}(u\cdot r)(u\cdot k)^{s-1} \right) \ , \nonumber
\end{align}
where $\calT^{(s)}(x,t,u)$ denotes the traceless part of $t^{\mu_1}\dots t^{\mu_s}$, as in \eqref{eq:T}.

Now, as discussed above, the curvature of $\Box_\ell\phi^{(s)}$ vanishes at $R\neq 0$. Therefore, the rational-function piece of \eqref{eq:Box_phi_s} at $R\neq 0$ must be a pure-gauge field $(u\cdot\nabla)\Theta^{(s)}$, i.e. the entire field $\Box_\ell\phi^{(s)}$ must be a sum of $(u\cdot\nabla)\Theta^{(s)}$ and a delta-function term. This is indeed the case, and the result reads (for $s\geq 1$):
\begin{align}
 \Box_\ell \phi^{(s)} = (u\cdot \nabla)\Theta^{(s)} + (i\sqrt{2})^s\left(\calT^{(s)}(x,t,u) - \frac{\theta(s-2)}{4s(s-1)}(g_{\mu\nu}u^\mu u^\nu)\calT^{(s-2)}(x,t,u)\right)\Box_\ell \phi^{(0)} \ , \label{eq:Box_phi_Theta}
\end{align}
with the gauge parameter:
\begin{align}
 \Theta^{(s)}(x,u;\ell,\ell') = -\frac{(i\sqrt 2)^{s-1}(1+iR)^2(u\cdot k)^{s-1}}{16\pi^2(s-1)!R^2(\ell\cdot x)^2\sqrt{-2\ell\cdot\ell'}} \ . \label{eq:Theta}
\end{align}
Note that for $s\geq 2$, the delta-function term in \eqref{eq:Box_phi_Theta} is different from that in \eqref{eq:Box_phi_s}. Moreover, while $\phi^{(s)}$ and therefore $\Box_\ell\phi^{(s)}$ are traceless, the delta-function term in \eqref{eq:Box_phi_Theta} is not: instead, its tensor structure is precisely that of the Fronsdal tensor $\calF\phi^{(s)}$ from \eqref{eq:Fronsdal_phi} (we will see later that this is not a coincidence). The discrepancy between the delta-function pieces in \eqref{eq:Box_phi_s} and \eqref{eq:Box_phi_Theta} implies that the pure-gauge term $(u\cdot \nabla)\Theta^{(s)}$ also contains a delta-function piece, and that this piece is not traceless (even though $(u\cdot \nabla)\Theta^{(s)}$ \emph{is} traceless for all $R\neq 0$).

The rest of this section is devoted to deriving eqs. \eqref{eq:Box_phi_0}-\eqref{eq:Theta}.

\subsection{Calculating the Laplacian at $R\neq 0$}

At $R\neq 0$, the spin-0 statement \eqref{eq:Box_phi_0} is simply that $\Box_\ell\phi^{(0)}$ vanishes. This follows immediately as a special case of \eqref{eq:Box_1_r}:
\begin{align}
 \Box_\ell\frac{1}{R\sqrt{-\ell\cdot\ell'}} = \Box_\ell\frac{1}{\sqrt{-\ell_\mu X^{\mu\nu}\ell'_\nu}} = 0 \ .
\end{align}
For general spins, deriving the rational term in \eqref{eq:Box_phi_s} is tedious, but straightforward. The relevant intermediate formulas read:
\begin{align}
 \begin{split}
   &\del^\mu_\ell\frac{1}{R\sqrt{-\ell\cdot\ell'}} = -\frac{\ell'^\mu + 2(\ell'\cdot x)x^\mu}{2(R\sqrt{-\ell\cdot\ell'})^3} = -\frac{1+R^2}{4R^3(\ell\cdot x)\sqrt{-\ell\cdot\ell'}}(x^\mu + t^\mu + r^\mu) \ ; \\
   &\del^\mu_\ell R = -\frac{(1+R^2)\ell'^\mu + 2(\ell'\cdot x)x^\mu}{2R(\ell\cdot\ell')} = \frac{1+R^2}{4R(\ell\cdot x)}\big[(1-R^2)x^\mu + (1+R^2)(t^\mu + r^\mu) \big] \ ; \\
   &\Box_\ell\frac{1}{R} = \frac{(\ell'\cdot x)^2}{R^3(\ell\cdot\ell')^2} = \frac{(1+R^2)^2}{4R^3(\ell\cdot x)^2} \ ; \\
   &\del^\mu_\ell r^\nu = -\del^\mu_\ell t^\nu = \frac{1}{2(\ell\cdot x)}\left(\eta^{\mu\nu} - \frac{x^\mu\ell^\nu}{\ell\cdot x} \right) = \frac{g^{\mu\nu} + x^\mu(t^\nu - r^\nu)}{2(\ell\cdot x)} \ ; \\
   &\Box_\ell r^\mu = -\Box_\ell t^\mu = \frac{t^\mu - r^\mu}{(\ell\cdot x)^2} \ ; \\
   &\del^\mu_\ell k^\nu = \frac{i(1+iR)}{4R(\ell\cdot x)}\left(g^{\mu\nu} + x^\mu(t^\nu - r^\nu) - \frac{1-iR}{2R^2}\big[(1-R^2)x^\mu + (1+R^2)(t^\mu + r^\mu) \big] r^\nu \right) \ ; \\ 
   &\Box_\ell k^\mu = \frac{i(1+iR)^2}{2R(\ell\cdot x)^2}\left(k^\mu - \frac{1+R^2}{4R^2}\,r^\mu \right) \ ; \\
   &\eta_{\nu\rho}\del^\nu_\ell\frac{1}{R\sqrt{-\ell\cdot\ell'}}\,\del_\ell^\rho k^\mu = -\frac{(1+iR)^2(1+R^2)}{8R^4(\ell\cdot x)^2\sqrt{-\ell\cdot\ell'}}\,r^\mu \ ; \\
   &\eta_{\rho\sigma}\del^\rho_\ell k^\mu\del^\sigma_\ell k^\nu = -\frac{(1+iR)^2}{16R^2(\ell\cdot x)^2} \left(g^{\mu\nu} - 4k^\mu k^\nu + \frac{4i(1+R^2)}{R}k^{(\mu}r^{\nu)} \right) \ ,
\end{split} \label{eq:ell_derivatives}
\end{align}
where we slightly abused notation by denoting $\Box_\ell \equiv \del_\ell\cdot\del_\ell$ for quantities that don't have conformal weight $\frac{1}{2}$ with respect to $\ell^\mu$. A key feature is the absence of an $r^\mu r^\nu$ term in the last line of \eqref{eq:ell_derivatives}.

\subsection{Guessing and verifying the gauge parameter}

The structure \eqref{eq:Theta} of the gauge parameter $\Theta^{(s)}$ is easy to guess, using the weights with respect to $\ell^\mu$ and $\ell'^\mu$, up to a coefficient of the form $f(R)$. We can then fix the function $f(R)$ up to a numerical factor, by demanding that $(u\cdot \nabla)\Theta^{(s)}$ should be traceless, i.e. that $\Theta^{(s)}$ should be divergence-free, at $R\neq 0$:
\begin{align}
 \begin{split}
   0 &= (\del_u\cdot\nabla)\frac{f(R)(u\cdot k)^{s-1}}{R^2(\ell\cdot x)^2} = \frac{(s-1)(u\cdot k)^{s-2}}{R^2}\,(k\cdot\nabla)\frac{f(R)}{(\ell\cdot x)^2} \\ 
     &= \frac{(s-1)(u\cdot k)^{s-2}}{R^2(\ell\cdot x)^2}\left(\frac{i}{2}f'(R) + \frac{f(R)}{1+iR} \right) \ , \label{eq:div_f}
 \end{split}
\end{align}
Here, in the second equality, we used the affine identity from \eqref{eq:k}, along with the useful formula (at $R\neq 0$):
\begin{align}
 \nabla_\mu\frac{k^\mu}{R^2} = 0 \ ,
\end{align}
while in the third equality, we used the bulk gradients:
\begin{align}
 \nabla_\mu(\ell\cdot x) = (\ell\cdot x)(r_\mu - t_\mu) \ ; \quad \nabla_\mu R = \frac{1+R^2}{R}\,r_\mu \ . \label{eq:gradients}
\end{align}
We can now read off from \eqref{eq:div_f} that $f(R)$ must take the form $\sim (1+iR)^2$, as in \eqref{eq:Theta}.

Having thus guessed $\Theta^{(s)}$ up to normalization, we can check our guess and normalize it by explicitly calculating the gradient $(u\cdot\nabla)\Theta^{(s)}$, and comparing with \eqref{eq:Box_phi_s} at $R\neq 0$. This is a straightforward calculation, using the bulk gradients in \eqref{eq:gradients} along with that of $k_\mu$ (at $R\neq 0$):
\begin{align}
 \nabla_\mu k_\nu = -\frac{2i}{R}\left(k_\mu k_\nu - \frac{1}{4}g_{\mu\nu} - \frac{i(1+R^2)}{R}k_{(\mu}r_{\nu)} \right) \ .
\end{align}

\subsection{Finding the delta-function contribution to $\Box_\ell\phi^{(s)}$} \label{sec:Box:full_delta}

Let us now derive the $\sim\delta^3(x;\ell,\ell')$ terms in eqs. \eqref{eq:Box_phi_0}-\eqref{eq:Box_phi_s}. In this derivation, we can truncate all expressions to just the leading terms in small $R$. Our strategy will be to fix $x^\mu$ and $\ell'^\mu$, and then integrate $\Box_\ell\phi^{(s)}$ over $\ell^\mu$ in an infinitesimal ball $B_3$ around $\ell^\mu = X^\mu{}_\nu\ell'^\nu$, i.e. around $R=0$. More precisely, we should integrate $\frac{1}{\sqrt{-\ell\cdot\ell'}}\Box_\ell\phi^{(s)}$, which has the appropriate weight $\Delta = 3$ with respect to $\ell^\mu$.

For the rational-function term in \eqref{eq:Box_phi_s}, this small-$R$ integral will vanish. Indeed, the integral of \emph{any} rational function of $R$ over an infinitesimal ball must either vanish or diverge: for powers $R^w$ with $w>-3$ the radial part of the integral vanishes, while for powers $w\leq -3$, it diverges; in the latter case, the whole integral may still vanish, if its angular part does. The rational function in \eqref{eq:Box_phi_s} has a dominant $\sim R^{-3}$ term, which implies that the integral may diverge. However, it cannot. The field $\phi^{(s)}$ is a well-defined distribution with respect to the boundary point $\ell^\mu$, and therefore so is $\Box_\ell\phi^{(s)}$. As a result, the integral of $\frac{1}{\sqrt{-2\ell\cdot\ell'}}\Box_\ell\phi^{(s)}$ over a compact region cannot diverge. Therefore, the angular integral of the $\sim R^{-3}$ piece must vanish, which one can indeed verify directly. We conclude that the rational term in \eqref{eq:Box_phi_s} does not contribute to the small-$R$ integral. Therefore, this integral is governed solely by the $\sim\delta^3(x;\ell,\ell')$ terms. Using the conversion \eqref{eq:delta_proportionality} between the bulk delta function $\delta^3(x;\ell,\ell')$ and the boundary delta function $\delta^{\frac{5}{2},\frac{1}{2}}(\ell,X\ell')$, the delta-function terms in \eqref{eq:Box_phi_s} then correspond to the statement:
\begin{align}
 \int_{B_3} d^3\ell\,\frac{\Box_\ell\phi^{(s)}}{\sqrt{-2\ell\cdot\ell'}} = -\frac{1}{2\pi} \times \left\{
  \begin{array}{cl}
     1 & \qquad s = 0 \\
     \displaystyle \frac{2(i\sqrt{2})^s}{s+1}\,\calT^{(s)}(x,t,u) & \qquad s\geq 1
  \end{array} \right. \ . \label{eq:Box_phi_integral}
\end{align}
To confirm that the integral \eqref{eq:Box_phi_integral} is indeed correct, we will use the Poincare parameterization \eqref{eq:varying_x_ell_1}-\eqref{eq:varying_x_ell_2}, with $x^\mu$ fixed to $(z,\mathbf{y})=(1,\vec 0)$, and $\ell^\mu(\mathbf{p})$ renamed for convenience into $\ell^\mu(-\mathbf{y})$:
\begin{align}
 x^\mu = (1,0,\vec{0}) \ ; \quad \ell^\mu = \left(\frac{1 + \mathbf{y}^2}{2}, \frac{1-\mathbf{y}^2}{2}, -\mathbf{y} \right) \ ; \quad \ell'^\mu = \left(\frac{1}{2}, -\frac{1}{2}, \vec{0} \right) \ . \label{eq:varying_ell}
\end{align}
The conformal Laplacian $\Box_\ell$ and the measure $d^3\ell$ become simply:
\begin{align}
 \Box_\ell = \del_{\mathbf{y}}\cdot\del_{\mathbf{y}} \ ; \quad d^3\ell = d^3\mathbf{y} \ .
\end{align}
The scalar products \eqref{eq:flat_scalar_products} in the small-$\mathbf{y}$ limit read:
\begin{align}
 \ell\cdot\ell' = \ell\cdot x = \ell'\cdot x = -\frac{1}{2} \ . \label{eq:flat_scalar_products_z_1}
\end{align} 
The other building blocks of $\phi^{(s)}$ become:
\begin{align}
 R = |\mathbf{y}| \ ; \quad t^\mu = (0,1,\vec 0) ; \quad r^\mu = (0,0,\mathbf{y}) \ ; \quad k^\mu = \frac{1}{2}\left(0,1,\frac{i\mathbf{y}}{|\mathbf{y}|} \right) \ . \label{eq:flat_R_k}
\end{align}
$\phi^{(s)}$ itself thus reads:
\begin{align}
 \phi^{(s)} = \frac{1}{4\sqrt{2}\pi^2 R} \times \left\{
   \begin{array}{cl}
     1 & \qquad s = 0 \\
     \displaystyle \frac{2}{s!}(i\sqrt{2})^s(u\cdot k)^s & \qquad s\geq 1
   \end{array} \right. \ . \label{eq:flat_phi}
\end{align}
To evaluate the integral \eqref{eq:Box_phi_integral}, we must now integrate $(\del_{\mathbf{y}}\cdot\del_{\mathbf{y}})\phi^{(s)}$ over a small ball around $\mathbf{y} = 0$, which can be done by integrating the radial derivative $\frac{\mathbf{y}}{|\mathbf{y}|}\cdot\del_{\mathbf{y}}\phi^{(s)}$ over a small 2-sphere of fixed $|\mathbf{y}|$. Conveniently, for $k^\mu$ given by \eqref{eq:flat_R_k}, we have $(\mathbf{y}\cdot\del_{\mathbf{y}})k^\mu = 0$. Thus, the derivative acts only on the $1/R$ factor in \eqref{eq:flat_phi}, giving:
\begin{align}
 (\mathbf{y}\cdot\del_{\mathbf{y}}) \phi^{(s)} = -\frac{1}{4\sqrt{2}\pi^2 |\mathbf{y}|} \times \left\{
    \begin{array}{cl}
      1 & \qquad s = 0 \\
      \displaystyle \frac{2}{s!}(i\sqrt{2})^s(u\cdot k)^s & \qquad s\geq 1
    \end{array} \right. \ .
\end{align}
We now need to divide this by $|\mathbf{y}|$ and integrate over the 2-sphere. This is a bit non-trivial for $s\geq 1$, where we must find the 2-sphere integral of $(u\cdot k)^s$. However, since $(u\cdot k)^s$ is traceless, the answer is constrained by symmetry to be a multiple of the traceless tensor $\calT^{(s)}(x,t,u)$ from \eqref{eq:T}. The proportionality coefficient can be fixed by evaluating both objects at $u^\mu = t^\mu$, i.e. taking all the tensor indices to point along $t^\mu$. We get:
\begin{align}
 \frac{1}{s!}\int_{S_2}d^2\Omega\,(u\cdot k)^s = \frac{4\pi}{s+1}\,\calT^{(s)}(x,t,u) \ , \label{eq:sphere_average}
\end{align}
which leads directly to the result \eqref{eq:Box_phi_integral}.

\subsection{Finding the delta-function contribution to $(u\cdot\nabla)\Theta^{(s)}$}

Let us now analyze the pure-gauge field $(u\cdot\nabla)\Theta^{(s)}$ near $R=0$. Recall that $(u\cdot\nabla)\Theta^{(s)}$ is the sum of a rational term (the same as in the full field $\Box_\ell\phi^{(s)}$), and a delta-function term (different from that in $\Box_\ell\phi^{(s)}$). To find the coefficient of the delta-function term, we again integrate over a small ball $B_3$ around $R=0$ (the rational term won't contribute, for the same reason as in section \ref{sec:Box:full_delta}). This time, the integral will be over $x^\mu$, with $\ell^\mu,\ell'^\mu$ held fixed. In particular, we can use the Poincare parameterization \eqref{eq:varying_x_ell_1}-\eqref{eq:varying_x_ell_2}, this time with $\ell^\mu$ fixed at $\mathbf{p} = 0$, and $x^\mu$ varying along the hypersurface $z=1$:
\begin{align}
 x^\mu = \left(1 + \frac{\mathbf{y}^2}{2}, \frac{\mathbf{y}^2}{2}, \mathbf{y} \right) \ ; \quad \ell^\mu = \left(\frac{1}{2}, \frac{1}{2}, \vec{0} \right) \ ; \quad \ell'^\mu = \left(\frac{1}{2}, -\frac{1}{2}, \vec{0} \right) \ . \label{eq:varying_x}
\end{align}
The 3d bulk hypersurface $x^\mu(\mathbf{y})$ intersects the $\gamma(\ell,\ell')$ geodesic perpendicularly at $\mathbf{y}=0$. The volume measure on the hypersurface is simply $d^3\mathbf{y}$. Thus, integral of $(u\cdot\nabla)\Theta^{(s)}$ over a small ball around $\mathbf{y}=0$ will directly give the coefficient of $\delta^3(x;\ell,\ell')$ in the delta-function piece of $(u\cdot\nabla)\Theta^{(s)}$.

At small $\mathbf{y}$, the building blocks of $\Theta^{(s)}$ again take the same form as in \eqref{eq:flat_scalar_products_z_1}-\eqref{eq:flat_R_k}. At small $R = |\mathbf{y}|$, the formula \eqref{eq:Theta} for $\Theta^{(s)}$ reads:
\begin{align}
 \Theta^{(s)} = -\frac{(i\sqrt{2})^{s-1}(u\cdot k)^{s-1}}{4\pi^2(s-1)!R^2} \ . \label{eq:Theta_small_R}
\end{align}
For $s=1$, this is just $\Theta^{(1)}\sim 1/R^2$, with gradient $(u\cdot\nabla)\Theta^{(1)}\sim (u\cdot r)/R^4$, whose integral over $B_3$ vanishes due to spherical symmetry. Let us now focus on $s\geq 2$. Before tackling the gradient $(u\cdot\nabla)\Theta^{(s)}$, consider first the small-$R$ integral of the \emph{divergence} $(\del_u\cdot\nabla)\Theta^{(s)}$. Via the Gauss theorem, this becomes a 2-sphere integral, which we then evaluate using \eqref{eq:sphere_average}:
\begin{align}
 \begin{split}
   \int_{B_3} d^3x\,(\del_u\cdot\nabla)\Theta^{(s)} &= R\int_{S_2}d^2\Omega\,(r\cdot\del_u)\Theta^{(s)} = -\frac{(i\sqrt{2})^s}{8\sqrt{2}\pi^2(s-2)!}\int d^2\Omega\,(u\cdot k)^{s-2} \\
     &= -\frac{(i\sqrt{2})^s}{2\sqrt{2}\pi(s-1)}\,\calT^{(s-2)}(x,t,u) \ .
 \end{split} \label{eq:div_Theta}
\end{align}
Now, let's return to the gradient $(u\cdot\nabla)\Theta^{(s)}$. As usual in the Fronsdal framework, since $\Theta^{(s)}$ is traceless, $(u\cdot\nabla)\Theta^{(s)}$ must be \emph{double-traceless}. Its small-$R$ integral is then constrained by symmetry to take the form:
\begin{align}
 \int_{B_3} d^3x\,(u\cdot\nabla)\Theta^{(s)} = a\,\calT^{(s)}(x,t,u) + b\,g_{\mu\nu}u^\mu u^\nu \calT^{(s-2)}(x,t,u) \ , \label{eq:grad_Theta_a_b}
\end{align}
with some numerical coefficients $a,b$. These can be fixed by imposing two conditions. First, the trace of \eqref{eq:grad_Theta_a_b} should be consistent with \eqref{eq:div_Theta}. Acting with $\frac{1}{2}(\del_u\cdot\del_u)$ on both sides of \eqref{eq:grad_Theta_a_b}, we get:
\begin{align}
 \int_{B_3} d^3x\,(\del_u\cdot\nabla)\Theta^{(s)} = 2sb\,\calT^{(s-2)}(x,t,u) \quad \Longrightarrow \quad b = -\frac{(i\sqrt{2})^s}{4\sqrt{2}\pi s(s-1)} \ .
\end{align}
Second, we notice that any non-vanishing small-$R$ integral of $(u\cdot\nabla)\Theta^{(s)}$ must involve the gradient acting in the $r^\mu$ direction: otherwise, the power of $1/R$ in \eqref{eq:Theta_small_R} is too low. Therefore, $(u\cdot\nabla)\Theta^{(s)}$ must vanish once we set $u^\mu = t^\mu$. This gives:
\begin{align}
 \frac{a(s+1)}{2^s s!} + \frac{b(s-1)}{2^{s-2}(s-2)!} = 0 \quad \Longrightarrow \quad a = -\frac{4s(s-1)^2}{s+1}\,b = \frac{(i\sqrt{2})^s(s-1)}{\pi\sqrt{2}(s+1)} \ .
\end{align}
The solution is thus:
\begin{align}
 \int_{B_3} d^3x\,(u\cdot\nabla)\Theta^{(s)} = \frac{(i\sqrt{2})^s}{\pi\sqrt{2}} \left(\frac{s-1}{s+1}\,\calT^{(s)}(x,t,u) - \frac{g_{\mu\nu}u^\mu u^\nu}{4s(s-1)}\,\calT^{(s-2)}(x,t,u) \right) \ .
\end{align}
This, then, is the coefficient of $\delta^3(x;\ell,\ell')$ in the delta-function piece of $(u\cdot\nabla)\Theta^{(s)}$. Restoring the weights w.r.t. $\ell^\mu,\ell'^\mu$ using the scalar products \eqref{eq:flat_scalar_products_z_1}, we obtain $(u\cdot\nabla)\Theta^{(s)}$ as:
\begin{align}
 \begin{split}
   &(u\cdot\nabla)\Theta^{(s)} = \text{rational part} + \frac{(i\sqrt{2})^s}{8\pi(\ell\cdot x)^2\sqrt{-\ell\cdot\ell'}}\,\delta^3(x;\ell,\ell') \\
   &\qquad\qquad \times\left(\frac{s-1}{s+1}\,\calT^{(s)}(x,t,u) - \frac{\theta(s-2)}{4s(s-1)}(g_{\mu\nu}u^\mu u^\nu)\calT^{(s-2)}(x,t,u) \right) \ .
 \end{split}
\end{align}
Combining this with \eqref{eq:Box_phi_s}, we obtain the delta-function terms in \eqref{eq:Box_phi_Theta}. This concludes the derivation of our formulas \eqref{eq:Box_phi_0}-\eqref{eq:Theta} for the boundary Laplacian $\Box_\ell\phi^{(s)}$.

\section{New gauges} \label{sec:Phi}

\subsection{Derivation and properties}

The result \eqref{eq:Box_phi_Theta} from the previous subsection has a clear implication: there exists a \emph{new, non-traceless gauge} for the Didenko-Vasiliev solution, such that its boundary Laplacian $\Box_\ell$ is given by just the delta-function terms in \eqref{eq:Box_phi_0},\eqref{eq:Box_phi_Theta}. Let us denote this new gauge by $\Phi^{(s)}$, as opposed to the original gauge $\phi^{(s)}$. We thus have:
\begin{align}
 \Box_\ell\Phi^{(s)} = -\frac{(i\sqrt{2})^s\delta^3(x;\ell,\ell')}{8\pi(\ell\cdot x)^2\sqrt{-\ell\cdot\ell'}} \left(\calT^{(s)}(x,t,u) - \frac{\theta(s-2)}{4s(s-1)}(g_{\mu\nu}u^\mu u^\nu)\calT^{(s-2)}(x,t,u) \right) \ . \label{eq:Box_Phi}
\end{align}
Similarly, there must exist another gauge $\Phi'^{(s)}$, with analogous behavior with respect to $\ell'$:
\begin{align}
 \Box_{\ell'}\Phi'^{(s)} = -\frac{(i\sqrt{2})^s\delta^3(x;\ell,\ell')}{8\pi(\ell'\cdot x)^2\sqrt{-\ell\cdot\ell'}} \left(\calT^{(s)}(x,t,u) - \frac{\theta(s-2)}{4s(s-1)}(g_{\mu\nu}u^\mu u^\nu)\calT^{(s-2)}(x,t,u) \right) \ . \label{eq:Box_Phi'} 
\end{align}
The fields $\Phi^{(s)},\Phi'^{(s)}$ themselves can be deduced from \eqref{eq:Box_Phi}-\eqref{eq:Box_Phi'} by inverting the boundary Laplacians, i.e. by integrating \eqref{eq:Box_Phi}-\eqref{eq:Box_Phi'} against the boundary propagator \eqref{eq:G_CFT}. Upon converting the geodesic delta function $\delta^3(x;\ell,\ell')$ into a boundary delta function $\delta^{\frac{5}{2},\frac{1}{2}}(\ell,X\ell')$ or $\delta^{\frac{1}{2},\frac{5}{2}}(X\ell,\ell')$ via \eqref{eq:delta_proportionality}, the integration becomes immediate. The answer reads:
\begin{align}
  \Phi^{(s)} &= \frac{(i\sqrt{2})^s}{8\pi^2 R\sqrt{-\ell\cdot\ell'}} \left(\calT^{(s)}(x,t_+,u) - \frac{\theta(s-2)}{4s(s-1)}(g_{\mu\nu}u^\mu u^\nu)\calT^{(s-2)}(x,t_+,u) \right) \ ; \label{eq:Phi} \\
  \Phi'^{(s)} &= \frac{(i\sqrt{2})^s}{8\pi^2 R\sqrt{-\ell\cdot\ell'}} \left(\calT^{(s)}(x,t_-,u) - \frac{\theta(s-2)}{4s(s-1)}(g_{\mu\nu}u^\mu u^\nu)\calT^{(s-2)}(x,t_-,u) \right) \ , \label{eq:Phi'}
\end{align} 
where $t_\pm^\mu \equiv t^\mu \pm r^\mu$. The $\sim \frac{1}{R\sqrt{-\ell\cdot\ell'}}$ scalar factor in \eqref{eq:Phi}-\eqref{eq:Phi'} is the same as in the old gauge \eqref{eq:phi}. As for the tensor structure of \eqref{eq:Phi}-\eqref{eq:Phi'}, it is inherited from \eqref{eq:Box_Phi}-\eqref{eq:Box_Phi'}, and coincides with that of the Fronsdal tensor \eqref{eq:Fronsdal_phi}, but with $t^\mu$ replaced by $t_\pm^\mu$. The significance of the vectors $t_\pm^\mu$ becomes clear by writing them as:
\begin{align}
 t_+^\mu(x;\ell') &= \frac{\ell'^\mu}{\ell'\cdot x} - \frac{x^\mu}{x\cdot x} = t^\mu(x;X\ell',\ell') \ ; \\
 t_-^\mu(x;\ell) &= \frac{x^\mu}{x\cdot x} - \frac{\ell^\mu}{\ell\cdot x} = t^\mu(x;\ell,X\ell) \ .
\end{align}
Thus, $t_+^\mu$ is the unit tangent at $x$ to the geodesic $\gamma(X\ell',\ell')$ that passes through $x$ and $\ell'$, while $t_-^\mu$ is the unit tangent at $x$ to the geodesic $\gamma(\ell,X\ell)$ that passes through $x$ and $\ell$. In particular, $t_\pm^\mu$ are unit vectors everywhere (whereas $t^\mu$ is unit only at $R=0$), and depend on only \emph{one} of the two boundary points $\ell,\ell'$. Under the interchange $\ell\leftrightarrow\ell'$, we have $t_\pm^\mu\rightarrow -t_\mp^\mu$. The two gauges \eqref{eq:Phi}-\eqref{eq:Phi'} are thus related via:
\begin{align}
 \Phi'^{(s)}(x,u;\ell,\ell') = (-1)^s\,\Phi^{(s)}(x,u;\ell',\ell) \ . \label{eq:Phi_symmetry}
\end{align}
A trivial check of the solutions \eqref{eq:Phi}-\eqref{eq:Phi'} is to re-apply the appropriate boundary Laplacian. This immediately reproduces e.g. \eqref{eq:Box_Phi} for $\Phi^{(s)}$: the $\Box_\ell$ will only act on the $\sim \frac{1}{R\sqrt{-\ell\cdot\ell'}}$ scalar prefactor, because the tensor structure constructed out of $t_+^\mu$ is $\ell$-independent. Another check is to directly evaluate the Fronsdal curvature \eqref{eq:Fronsdal} of e.g. $\Phi^{(s)}$, and show that it agrees with the result \eqref{eq:Fronsdal_phi} for $\phi^{(s)}$. This calculation is detailed in section \ref{sec:Phi:bulk_check} below. A notable intermediate result is the de Donder tensor \eqref{eq:deDonder} of $\Phi^{(s)}$, which reads (for $s\geq 1$):
\begin{align}
 \calD\Phi^{(s)} = \frac{(i\sqrt{2})^s}{4\pi^2 R\sqrt{-\ell\cdot\ell'}}\,\calT^{(s-1)}(x,t_+,u) \ . \label{eq:deDonder_Phi}
\end{align}
To our knowledge, the solutions \eqref{eq:Phi}-\eqref{eq:Phi'} constitute a pair of \emph{hitherto unknown gauges} for the linearized HS fields of a point-particle source in (A)dS spacetime. This appears to be true even for $s=2$, i.e. for linearized General Relativity on an (A)dS background. What's unusual about these new gauges is that they satisfy none of the simple bulk gauge conditions -- they are neither traceless, nor transverse, nor de Donder -- but instead they have a simple relationship with \emph{one of the two boundary endpoints} of the source particle's worldline.

It is interesting to consider the $R\ll 1$ limit of \eqref{eq:Phi}-\eqref{eq:Phi'}. There, $t_\pm^\mu$ can be approximated as $t^\mu$, so the two gauges $\Phi^{(s)},\Phi'^{(s)}$ become a single solution $\Phi_{\bbR^4}^{(s)}$. Since the curvature of $EAdS_4$ is negligible at $R\ll 1$, this is a solution for the linearized HS fields of a point-particle source \emph{in flat spacetime $\bbR^4$}:
\begin{align}
 \Phi_{\bbR^4}^{(s)} \sim \frac{(i\sqrt{2})^s}{R} \left(\calT^{(s)}(t,u) - \frac{\theta(s-2)}{4s(s-1)}(u\cdot u)\calT^{(s-2)}(t,u) \right) \ , \label{eq:Phi_flat_limit}
\end{align} 
where all vectors are now 4-dimensional, $R$ is the flat distance from the source geodesic, $t^\mu$ is the unit vector parallel to it, and $\calT^{(s)}(t,u)$ is the traceless part of $t^{\mu_1}\dots t^{\mu_s}$. Unlike the $EAdS_4$ solutions \eqref{eq:Phi}, their flat limit \eqref{eq:Phi_flat_limit} satisfies the de Donder gauge condition:
\begin{align}
 \calD\Phi_{\bbR^4}^{(s)} \equiv \left( \del_u\cdot\del_x  - \frac{1}{2}(u\cdot\del_x)(\del_u\cdot\del_u) \right)\Phi_{\bbR^4}^{(s)} = 0 \ . \label{eq:deDonder_flat}
\end{align}
This is consistent with the full $EAdS_4$ result \eqref{eq:deDonder_Phi}: by dimensional analysis, a non-vanishing derivative of a $\sim 1/R$ field in flat spacetime would be $\sim 1/R^2$, which is much larger than the $\sim 1/R$ result in \eqref{eq:deDonder_Phi}.

\subsection{Calculating bulk derivatives of $\Phi^{(s)}$} \label{sec:Phi:bulk_check}

Here, we detail the bulk calculation of the de Donder tensor \eqref{eq:deDonder_Phi} and Fronsdal tensor \eqref{eq:Fronsdal_phi} in the new gauge $\Phi^{(s)}$. The calculation's basic ingredients are $\nabla_\mu R$ from \eqref{eq:gradients}, along with:
\begin{align}
 \nabla^\mu t^\nu_+ = q_+^{\mu\nu} \ ; \quad \nabla^\mu q_+^{\nu\rho} = -2 q_+^{\mu(\nu} t_+^{\rho)} \ ; \quad r\cdot t_+ = \frac{R^2}{1+R^2} \ ; \quad r_\nu q_+^{\nu\mu} = r^\mu - \frac{R^2}{1+R^2}\,t_+^\mu \ ,
\end{align}
where $q_+^{\mu\nu} \equiv g^{\mu\nu} - t_+^\mu t_+^\nu$ is the 3d metric in perpendicular to $t_+^\mu$. 

Let us now confirm the expression \eqref{eq:deDonder_Phi} for the de Donder tensor $\calD\Phi^{(s)}$, recalling that it's simply the traceless part of the divergence $(\del_u\cdot\nabla)\Phi^{(s)}$. We begin by packaging the expression \eqref{eq:T} for the traceless structure $\calT^{(p)}(x,t_+,u)$ as:
\begin{align}
 \calT^{(p)}(x,t_+,u) = \frac{1}{p!}\left((t_+\cdot u)^p - \frac{p-1}{4}\,\theta(p-2)(g_{\mu\nu}u^\mu u^\nu) (t_+\cdot u)^{p-2}\right) - \text{double traces} \ . \label{eq:T_expansion}
\end{align}
Upon contracting with $\nabla_\mu R\sim r_\mu$, this becomes:
\begin{align}
 &(r\cdot\del_u)\calT^{(p)}(x,t_+,u) = \frac{1}{p!}\left(\pm \frac{pR^2}{1+R^2}(t_+\cdot u)^{p-1} - \frac{p-1}{2}\,\theta(p-2)(r\cdot u)(t_+\cdot u)^{p-2} \right) - \text{traces} \nonumber \\
 &\qquad = \pm\frac{R^2}{1+R^2}\,\calT^{(p-1)}(x,t_+,u) - \frac{\theta(p-2)}{2p}(r\cdot u)\calT^{(p-2)}(x,t_+,u) - \text{traces} \ . \label{eq:T_r_contraction}
\end{align}
The gradient of \eqref{eq:T_expansion} reads:
\begin{align}
 \nabla^\mu\calT^{(p)}(x,t_+,u) ={}& \frac{q_+^{\mu\nu}u_\nu}{p!}\left(p(t_+\cdot u)^{p-1} - \frac{(p-1)(p-2)}{4}\,\theta(p-2)(g_{\rho\sigma}u^\rho u^\sigma)(t_+\cdot u)^{p-3} \right) \nonumber \\
    &- \text{double traces} \ .
\end{align}
Contracting with $\del_u^\mu$ and $u^\mu$ respectively, we obtain:
\begin{align}
 (\del_u\cdot\nabla)\calT^{(p)}(x,t_+,u) &= \frac{(p+1)(p+2)}{2p}\,\theta(p-1)\calT^{(p-1)}(x,t_+,u) \ ; \label{eq:T_divergence} \\
 (u\cdot\nabla)\calT^{(p)}(x,t_+,u) &= \pm\left(-p(p+1)\calT^{(p+1)}(x,t_+,u) + \frac{p+2}{4p}\,\theta(p-1)(g_{\mu\nu}u^\mu u^\nu)\calT^{(p-1)}(x,t_+,u) \right) \ . \label{eq:T_gradient}
\end{align}
Putting eqs. \eqref{eq:T_r_contraction},\eqref{eq:T_divergence}-\eqref{eq:T_gradient} together, the result \eqref{eq:deDonder_Phi} for the de Donder tensor follows. Crucially, the $\sim r\cdot u$ pieces cancel between the contributions from $(\nabla\frac{1}{R}\cdot\del_u)\calT^{(s)}$ and from $(\nabla\frac{1}{R}\cdot u)\calT^{(s-2)}$. It is this cancellation that is responsible for the vanishing \eqref{eq:deDonder_flat} of the de Donder tensor in the flat limit.

Let's now plug in the result \eqref{eq:deDonder_Phi} for $\calD\Phi^{(s)}$ to calculate the Fronsdal tensor using the formula \eqref{eq:Fronsdal_deDonder}. At $R\neq 0$, the additional formulas required for this calculation read:
\begin{align}
 (\nabla\cdot\nabla)\calT^{(p)}(x,t_+,u) &= -p(p+2)\calT^{(p)}(x,t_+,u) \ ; \\ 
 (r\cdot\nabla)\calT^{(p)}(x,t_+,u) &= -\frac{pR^2}{1+R^2}\calT^{(p)}(x,t_+,u) + \theta(p-1)(r\cdot u)\calT^{(p-1)}(x,t_+,u) - \text{traces} \ .
\end{align}
In the latter expression, we will need to expand one more level of traces, via:
\begin{align}
 \begin{split}
   &(r\cdot u)\calT^{(p-1)}(x,t_+,u) - \text{traces} = (r\cdot u)\calT^{(p-1)}(x,t_+,u) \\
     &\quad{}- \frac{g_{\mu\nu}u^\mu u^\nu}{2p}\left( \frac{\theta(p-2)R^2}{1+R^2}\,\calT^{(p-2)}(x,t_+,u) - \frac{\theta(p-3)}{2(p-1)}(r\cdot u)\calT^{(p-3)}(x,t_+,u) \right) \\
     &\quad{}- \text{double traces} \ . 
 \end{split}
\end{align}
Putting all of this together, we obtain that the Fronsdal tensor $\calF\Phi^{(s)}$ vanishes at $R\neq 0$ up to double traces. Since it is in any case double-traceless, this implies that it simply vanishes at $R\neq 0$.

Finally, let us consider the delta-function contribution to the Fronsdal tensor at $R=0$. To compute it, it's sufficient to work in the $R\ll 1$ flat limit \eqref{eq:Phi_flat_limit}, where the de Donder tensor $\calD\Phi^{(s)}$ vanishes. Therefore, a delta-function contribution can only come from the $\nabla\cdot\nabla$ term in \eqref{eq:Fronsdal_deDonder}. As usual, we can find this by integrating $(\frac{r}{R}\cdot\nabla)\Phi^{(s)}$ over an infinitesimal 2-sphere. Here, the only non-vanishing contribution is obtained when the derivative acts on the $1/R$ factor in \eqref{eq:Phi}: if the derivative acts instead on the $\calT^{(s)},\calT^{(s-2)}$ tensor structures, the result will not have a sufficiently negative power of $R$, and in any case will vanish upon angular averaging. Thus, the delta-function contribution to the Fronsdal tensor is actively produced solely by the $1/R$ factor, while the tensor structure is merely ``along for the ride''. The result \eqref{eq:Fronsdal_phi} for the Fronsdal tensor now immediately follows. In addition, we now understand why the tensor structure in \eqref{eq:Box_phi_Theta}, which propagated into \eqref{eq:Phi}, is the same as in \eqref{eq:Fronsdal_phi}.

\section{Analytic bulk derivation of bilocal-bilocal correlator} \label{sec:analytic_2_point}

Armed with the new gauges of section \ref{sec:Phi}, we can now provide an analytic derivation of eqs. \eqref{eq:bilocal_2_point_raw}-\eqref{eq:bilocal_2_point}, which express the duality between the boundary bilocal correlator $\left<\calO(\ell,\ell')\calO(L,L)\right>$ and the mutual bulk action of two Didenko-Vasiliev solutions $\phi^{(s)}(x,u;\ell,\ell'),\phi^{(s)}(x,u;L,L')$. Let us denote the LHS of \eqref{eq:bilocal_2_point}, i.e. the mutual bulk action multiplied by $-2\pi\sqrt{-L\cdot L'}$, as:
\begin{align}
  W^\phi(\ell,\ell';L,L') \equiv \int_{-\infty}^\infty d\tau \sum_{s=0}^\infty s!(i\sqrt{2})^s \phi^{(s)}\big(x(\tau;L,L'),\dot x(\tau;L,L');\ell,\ell'\big) \ . \label{eq:W_phi}
\end{align}
Our task is then to prove that this equals the RHS of \eqref{eq:bilocal_2_point}. Our strategy will be as follows:
\begin{enumerate}
 \item Calculate the analogous action $W^\Phi$ in the new gauge $\Phi^{(s)}$, and show that it \emph{vanishes} (of course, a similar result holds also in the $\Phi'^{(s)}$ gauge).
 \item Show that upon acting with a boundary Laplacian $\Box_\ell$ or $\Box_{\ell'}$, the effect $W^\phi-W^\Phi$ of the gauge transformation \emph{also} vanishes, except possibly at coincident points.
 \item Show that $W^\phi$ is regular at $\ell=\ell'$, so that the Laplacians $\Box_\ell,\Box_{\ell'}$ must vanish there as well.
 \item Notice that this, together with the result of section \ref{sec:local_limit:action}, completely fixes the Taylor series of $W^\phi$ around $\ell=\ell'$, and therefore the function itself.
\end{enumerate}
Consider, then, replacing $\phi^{(s)}$ in \eqref{eq:W_phi} by the new gauge $\Phi^{(s)}$:
\begin{align}
 W^\Phi(\ell,\ell';L,L') = \int_{-\infty}^\infty d\tau \sum_{s=0}^\infty s!(i\sqrt{2})^s \Phi^{(s)}\big(x(\tau;L,L'),\dot x(\tau;L,L');\ell,\ell'\big) \ . \label{eq:W_Phi}
\end{align}
Since $\dot x^\mu$ is a unit vector, the $\Phi^{(s)}$ field in \eqref{eq:W_Phi} reads simply:
\begin{align}
 \Phi^{(s)}(x,\dot x;\ell,\ell') = -\frac{(i\sqrt{2})^s}{8\pi^2 R\sqrt{-\ell\cdot\ell'}} \left(\calT^{(s)}(x,t_+,\dot x) - \frac{\theta(s-2)}{4s(s-1)}\,\calT^{(s-2)}(x,t_+,\dot x) \right) \ , \label{eq:Phi_x_dot}
\end{align}
where $R$ and $t_+^\mu$ are defined in terms of $(x,\ell,\ell')$. Plugging this into \eqref{eq:W_Phi}, we see immediately that the contribution of the traceless piece $\sim\calT^{(s)}$ for each spin $s$ cancels against that of the trace piece $\sim\calT^{(s^*-2)}$ for spin $s^* = s+2$. In other words, \emph{due to the sum over spins}, and regardless of the $d\tau$ integral, the action in the $\Phi^{(s)}$ gauge vanishes:
\begin{align}
 W^\Phi(\ell,\ell';L,L') = 0 \ .
\end{align}
This cancellation between spins arises from an interplay of the $\sim (i\sqrt{2})^s$ spin-dependence of the charges in the Didenko-Vasiliev solution \eqref{eq:phi}-\eqref{eq:Einstein_phi} (which propagates into \eqref{eq:W_Phi}-\eqref{eq:Phi_x_dot}), and the factor of $\frac{1}{4}$ between the traceless and trace terms in \eqref{eq:Fronsdal_phi},\eqref{eq:Box_Phi},\eqref{eq:Phi} (which propagates into \eqref{eq:Phi_x_dot}). This makes the second known example of inter-spin cancellations in the interaction of Didenko-Vasiliev solutions; the first such example was the cancellation of UV divergences in \cite{David:2020fea}. See also \cite{Giombi:2013fka} for inter-spin cancellations in the HS multiplet in a different context.

Next, let us quantify the difference between the vanishing action \eqref{eq:W_Phi} and the desired one \eqref{eq:W_phi}. We do not know directly the gauge parameter that relates $\phi^{(s)}$ to $\Phi^{(s)}$, but we do know its $\Box_\ell$ boundary Laplacian:
\begin{align}
 \Box_\ell\!\left(\phi^{(s)} - \Phi^{(s)} \right) = (u\cdot\nabla)\Theta^{(s)} \ ,
\end{align} 
where $\Theta^{(s)}$ is the gauge parameter from \eqref{eq:Theta}. We thus have:
\begin{align}
 \Box_\ell W^\phi = \Box_\ell\!\left(W^\phi - W^\Phi \right) = \int_{-\infty}^\infty d\tau \sum_{s=1}^\infty s!(i\sqrt{2})^s (\dot x\cdot\nabla)\Theta^{(s)}\big(x(\tau;L,L'),\dot x(\tau;L,L');\ell,\ell'\big) \ .
\end{align}
Since $\dot x^\mu$ satisfies the geodesic equation $(\dot x\cdot \nabla)\dot x^\mu = 0$, the integrand is (as expected) a total derivative, yielding a boundary term:
\begin{align}
 \begin{split}
   \Box_\ell W^\phi &= \int_{-\infty}^\infty d\tau \sum_{s=1}^\infty s!(i\sqrt{2})^s \frac{d}{d\tau}\Theta^{(s)}\big(x(\tau;L,L'),\dot x(\tau;L,L');\ell,\ell'\big) \\
     &= \sum_{s=1}^\infty s!(i\sqrt{2})^s\left.\!\Theta^{(s)}\big(x(\tau;L,L'),\dot x(\tau;L,L');\ell,\ell'\big)\right|_{\tau=-\infty}^\infty \ .
 \end{split} \label{eq:Box_W}
\end{align}
Now, for non-coincident $(\ell,\ell',L,L')$, it's easy to see that $\Theta^{(s)}(x,\dot x;\ell,\ell')$ vanishes at both endpoints $\tau=\pm\infty$, for every spin $s$. Indeed, let us expand \eqref{eq:Theta} as:
\begin{align}
 \Theta^{(s)}\big(x,\dot x;\ell,\ell'\big) = -\frac{(1+iR)^2}{16\pi^2(s-1)!R^2(\ell\cdot x)^2\sqrt{-2\ell\cdot\ell'}}\left(\frac{i}{\sqrt{2}}\right)^{s-1}\left(t\cdot\dot x + \frac{ir\cdot\dot x}{R}\right)^{s-1} \ . \label{eq:Theta_x_dot}
\end{align}
The endpoints $x^\mu\sim e^{|\tau|}L^\mu$ and $x^\mu\sim e^{|\tau|}L'^\mu$ of the $\gamma(L,L')$ geodesic (with $\tau=\pm\infty$ respectively) are both in the large-$R$ regime with respect to $\gamma(\ell,\ell')$. Thus, the $\frac{(1+iR)^2}{R^2}$ factor in \eqref{eq:Theta_x_dot} tends to $-1$. Futhermore, in the large-$R$ regime, $t^\mu$ and $r^\mu/R$ become unit vectors. Since $\dot x$ is also a unit vector, this implies that the scalar products $t\cdot\dot x$ and $\frac{r\cdot\dot x}{R}$ are no larger than 1 in absolute value. Meanwhile, the $(\ell\cdot x)^2$ factor in the denominator of \eqref{eq:Theta_x_dot} tends to infinity, showing that the whole expression \eqref{eq:Theta_x_dot} vanishes. We conclude that $\Box_\ell W^\phi$ in \eqref{eq:Box_W} vanishes as well, with the possible exception of singularities at coincident points. This is of course consistent with the behavior of the boundary correlator on the RHS of \eqref{eq:bilocal_2_point}.

Now, coincident points don't \emph{necessarily} imply a singularity in $W^\phi$. In particular, we can show that the $\ell=\ell'$ limit is perfectly regular, again in agreement with \eqref{eq:bilocal_2_point}. Consider again the building blocks of $\phi^{(s)}(x,u;\ell,\ell')$, as expressed in \eqref{eq:phi_t_r}. Writing them out explicitly in terms of $(x^\mu,\ell^\mu,\ell'^\mu)$, we see that $\frac{1}{R\sqrt{-\ell\cdot\ell'}}$, $t^\mu$ and $r^\mu$ are all analytic at $\ell=\ell'$. Furthermore, $\frac{1}{R}$ is not analytic, being proportional to $\sqrt{-\ell\cdot\ell'}$, but $\frac{1}{R^2}$ \emph{is} analytic. Thus, the only non-analytic part of $\phi^{(s)}(x,u;\ell,\ell')$ is the one with odd powers of $\frac{i(r\cdot u)}{R}$ in \eqref{eq:phi_t_r}, i.e. the imaginary/real part of $\phi^{(s)}(x,u;\ell,\ell')$ for even/odd $s$. As mentioned above in section \ref{sec:preliminaries:DV_mutual}, it's easy to see that this part gives a vanishing contribution to $W^\phi$, for each spin $s$. Indeed, the contributions to the $d\tau$ integral from opposite sides of the geodesics' point of closest approach cancel, because they have equal values of $R$ and $t\cdot\dot x$, but equal and opposite values of $r\cdot\dot x$. Thus, $W^\phi$ receives contributions only from the part of $\phi^{(s)}(x,u;\ell,\ell')$ that is analytic at $\ell=\ell'$, and is therefore itself analytic there.

The upshot is that $W^\phi$ must satisfy the Laplace equation $\Box_\ell W^\phi = 0$ also at $\ell=\ell'$. Completely analogously, it must also satisfy $\Box_{\ell'} W^\phi = 0$ there. Now, consider the Taylor expansion of $W^\phi$ as a function of $(\ell,\ell')$ around a coincident point $\ell=\ell'$. This Taylor expansion is exhausted by:
\begin{itemize}
	\item The derivatives $D^{(s)}$ from \eqref{eq:D}, which construct local currents from bilocals. As we saw in section \ref{sec:local_limit:action}, these derivatives of $W^\phi$ at $\ell=\ell'$ are consistent with the desired holographic result \eqref{eq:bilocal_2_point}.
	\item ``Descendants'' of the above, i.e. further $\ell$ derivatives taken after imposing $\ell=\ell'$. These are automatically also consistent with \eqref{eq:bilocal_2_point}.
	\item Laplacians $\Box_\ell$ and $\Box_{\ell'}$, which vanish as we just saw, again consistently with \eqref{eq:bilocal_2_point}.
\end{itemize}
We conclude that the entire Taylor series of $W^\phi$, and therefore $W^\phi$ itself, is consistent with the desired holographic relation \eqref{eq:bilocal_2_point}.

\section{Double Laplacian} \label{sec:double_Box}

For our final result, it is interesting to also evaluate the \emph{double Laplacian} $\Box_\ell\Box_{\ell'}$ of the Didenko-Vasiliev solution: this will be the bulk dual of applying the boundary field equation to \emph{both} legs of the boundary bilocal $\calO(\ell,\ell')$. We will evaluate this double Laplacian in the new gauges \eqref{eq:Phi}-\eqref{eq:Phi'}. The double Laplacian of e.g. $\Phi'^{(s)}$ can be found by applying $\Box_\ell$ to our expression for $\Box_{\ell'}\Phi'^{(s)}$ in \eqref{eq:Box_Phi'}. To facilitate this calculation, we rewrite \eqref{eq:Box_Phi'} in terms of $t_+^\mu$ instead of $t^\mu$, using the fact that these are equal at $R=0$:
\begin{align}
 \Box_{\ell'}\Phi'^{(s)} = -\frac{(i\sqrt{2})^s\delta^3(x;\ell,\ell')}{8\pi(\ell'\cdot x)^2\sqrt{-\ell\cdot\ell'}} \left(\calT^{(s)}(x,t_+,u) - \frac{\theta(s-2)}{4s(s-1)}(g_{\mu\nu}u^\mu u^\nu)\calT^{(s-2)}(x,t_+,u) \right) \ . \label{eq:Box_Phi'_t_plus} 
\end{align}
The benefit of this is that the factor $\delta^3(x;\ell,\ell')/\sqrt{-\ell\cdot\ell'}$ is now the only part of \eqref{eq:Box_Phi'_t_plus} that depends on $\ell^\mu$. To this factor, we can now easily apply $\Box_\ell$, using eqs. \eqref{eq:delta_proportionality}-\eqref{eq:Box_delta_proportionality} to convert between boundary and bulk delta functions and their Laplacians:
\begin{align}
 \Box_\ell\frac{\delta^3(x;\ell,\ell')}{\sqrt{-\ell\cdot\ell'}} = 4\sqrt{2}(\ell'\cdot x)^2\,\Box_\ell\delta^{\frac{1}{2},\frac{5}{2}}(\ell,X\ell') = \frac{(\ell'\cdot x)^2}{(-\ell\cdot\ell')^{5/2}}(\nabla\cdot\nabla)\delta^3(x;\ell,\ell') \ .
\end{align}
Plugging this to apply $\Box_\ell$ to \eqref{eq:Box_Phi'_t_plus}, and performing the analogous procedure for $\Phi^{(s)}$, we obtain:
\begin{align}
 \begin{split}
   \Box_\ell\Box_{\ell'}\Phi^{(s)} ={}& {-}\frac{(i\sqrt{2})^s}{8\pi(-\ell\cdot\ell')^{5/2}}(\nabla\cdot\nabla)\delta^3(x;\ell,\ell') \\  
    &\times \left(\calT^{(s)}(x,t_-,u) - \frac{\theta(s-2)}{4s(s-1)}(g_{\mu\nu}u^\mu u^\nu)\calT^{(s-2)}(x,t_-,u) \right) \ ;
 \end{split} \label{eq:Double_Box_Phi_raw} \\
 \begin{split}
   \Box_\ell\Box_{\ell'}\Phi'^{(s)} ={}& {-}\frac{(i\sqrt{2})^s}{8\pi(-\ell\cdot\ell')^{5/2}}(\nabla\cdot\nabla)\delta^3(x;\ell,\ell') \\
    &\times \left(\calT^{(s)}(x,t_+,u) - \frac{\theta(s-2)}{4s(s-1)}(g_{\mu\nu}u^\mu u^\nu)\calT^{(s-2)}(x,t_+,u) \right) \ . 
 \end{split} \label{eq:Double_Box_Phi'_raw} 
\end{align}
With some further work, we can rewrite these expressions in terms of $t^\mu$ rather than $t_\pm^\mu$. More specifically, the challenge is to rewrite factors of $\calT^{(p)}(x,t_\pm,u)(\nabla\cdot\nabla)\delta^3(x;\ell,\ell')$ in terms of $t^\mu$. Since we're guaranteed that this expression is traceless, we can work up to traces. Thus, we need to rewrite in terms of $t^\mu$ the expression $\frac{1}{p!}(u\cdot t_\pm)^p(\nabla\cdot\nabla)\delta^3(x;\ell,\ell')$, up to trace terms. To accomplish this, we expand the factors of $t_\pm^\mu = t^\mu \pm r^\mu$, as:
\begin{align}
 (t_\pm\cdot u)^p = (t\cdot u)^p \pm p(r\cdot u)(t\cdot u)^{p-1} + \frac{p(p-1)}{2}(r\cdot u)^2(t\cdot u)^{p-2} + O(R^3) \ .
\end{align}
The resulting powers of $r^\mu$ can be handled through the ``integration by parts'' formulas:
\begin{align}
 (u\cdot r)^m(\nabla\cdot\nabla)\delta^3(x;\ell,\ell') = \left\{
   \begin{array}{cl}
     (\nabla\cdot\nabla)\delta^3(x;\ell,\ell') & \qquad m=0 \\
     -2(u\cdot\nabla)\delta^3(x;\ell,\ell') & \qquad m=1 \\
     2(q_{\mu\nu}u^\mu u^\nu)\delta^3(x;\ell,\ell') & \qquad m=2 \\
     0 & \qquad m \geq 3
  \end{array} \right. \ , \label{eq:delta_by_parts}
\end{align}
which follow from $\nabla_\mu r_\nu = q_{\mu\nu} + O(R^2)$. In the $m=2$ case, we can then expand $q_{\mu\nu} = g_{\mu\nu} - t_\mu t_\nu$, and discard the $g_{\mu\nu}$ as a trace piece. We end up with:
\begin{align}
 \begin{split}
   \calT^{(p)}(x,t_\pm,u)(\nabla\cdot\nabla)\delta^3(x;\ell,\ell') &= \calT^{(p)}(x,t,u)(\nabla\cdot\nabla)\delta^3(x;\ell,\ell') \\
     &\mp 2\theta(p-1)\left(\calT^{(p-1)}(x,t,u)(u\cdot\nabla)\delta^3(x;\ell,\ell') - \text{traces} \right) \\
     &- p(p-1)\,\calT^{(p)}(x,t,u)\,\delta^3(x;\ell,\ell') \ . \label{eq:T_t_pm_to_t}
 \end{split}
\end{align}
The second term in \eqref{eq:T_t_pm_to_t} is cumbersome, but it can be canceled from \eqref{eq:Double_Box_Phi_raw}-\eqref{eq:Double_Box_Phi'_raw} by switching to a symmetrized gauge:
\begin{align}
 \Phi_{\text{symm}}^{(s)} = \frac{1}{2}\left(\Phi^{(s)} + \Phi'^{(s)}\right) \ ,
\end{align}
which is equivalent to symmetrizing/antisymmetrizing over $\ell\leftrightarrow\ell'$ for even/odd $s$, respectively. The double Laplacian then takes the form:
\begin{align}
 \begin{split}
   &\Box_\ell\Box_{\ell'}\Phi_{\text{symm}}^{(s)} = -\frac{(i\sqrt{2})^s}{8\pi(-\ell\cdot\ell')^{5/2}} \left[ \vphantom{\frac{\theta(s-2)}{4s(s-1)}} \calT^{(s)}(x,t,u)\big(\nabla\cdot\nabla - s(s-1) \big)\delta^3(x;\ell,\ell') \right. \\
   &\qquad \left.{}- \frac{\theta(s-2)}{4s(s-1)}(g_{\mu\nu}u^\mu u^\nu)\calT^{(s-2)}(x,t,u)\big(\nabla\cdot\nabla - (s-2)(s-3) \big)\delta^3(x;\ell,\ell') \right] \ . \label{eq:Double_Box_Phi} 
 \end{split}
\end{align}

\section*{Acknowledgements}

We are grateful to Sudip Ghosh and Mirian Tsulaia for discussions. This work was supported by the Quantum Gravity Unit of the Okinawa Institute of Science and Technology Graduate University (OIST).

\end{document}